\documentclass[11pt, spanish]{article}
\usepackage{cite}
\usepackage{amsmath,amsfonts,amssymb}
\usepackage[small,bf,hang]{caption}
\usepackage{slashed}
\usepackage{color}
\usepackage{soul}
\usepackage{extarrows}
\usepackage{hyperref}

\usepackage{ulem}

\usepackage{geometry}
\def\hybrid{
        \topmargin -20pt
        \oddsidemargin 0pt
        \headheight 0pt \headsep 0pt
        \textwidth 6.25in 
        \textheight 9.5in 
        \marginparwidth .875in
        \parskip 5pt plus 1pt \jot = 1.5ex}

\hybrid

 \csname
@addtoreset\endcsname{equation}{section}

\usepackage{epsfig}
\usepackage{color}
\usepackage{graphics}
\usepackage{graphicx}
\usepackage{slashed}

\usepackage{amsthm}
\usepackage{amssymb}
\usepackage{amsmath}
\usepackage{amsfonts}

\numberwithin{equation}{section}

\newcommand{\be}{\begin{equation}}
\newcommand{\ee}{\end{equation}}
\newcommand{\ba}{\begin{eqnarray}}
\newcommand{\ea}{\end{eqnarray}}

\newcommand{\nn}{\nonumber}

\input xy
\xyoption{all}

\begin{document}
\begin{titlepage}
\rightline{}
\begin{center}
\vskip 1.5cm
 {\Large \bf{ Tri-vector symmetry of 11 dimensional supergravity}}
\vskip 0,7cm  

 {\large {Walter H. Baron$^1$, Diego Marqu\'es$^{2,3} $, Carmen A. N\'u\~nez$^2$ \\ 
 \vskip 0,6cm
 and Nahuel Yazbek$^{4}$}}
\vskip 1cm

$^1$ {\it  Instituto de F\'isica de La Plata}, (CONICET-UNLP)\\
{\it Diagoinal 113 /63 y 64, s/n, B1900, La Plata, Argentina}\\
\medskip

{\it Departamento de F\'isica, Universidad Nacional de La Plata. }\\
{\it C. 115 s/n, B1900, La Plata, Argentina}\\
wbaron@fisica.unlp.edu.ar,\\
 
\vskip .3cm

$^2$ {\it   Instituto de Astronom\'ia y F\'isica del Espacio}, (CONICET-UBA)\\
{\it Ciudad Universitaria, Pabell\'on IAFE, CABA, C1428ZAA, Argentina}\\
diegomarques@iafe.uba.ar, \ \ carmen@iafe.uba.ar\\

\vskip .3cm

$^3$ {\it Departamento de F\'isica, Universidad de Buenos Aires.}\\
{\it Ciudad Universitaria, Pabell\'on 1, CABA, C1428ZAA, Argentina}\\

\vskip .3cm

$^4$ {\it   Instituto de Matemática Luis Santaló}, (CONICET-UBA)\\
{\it Ciudad Universitaria, Pabell\'on I, CABA, C1428ZAA, Argentina}\\
nyazbek@dm.uba.ar, 
\vskip .1cm

\vskip .4cm

\vskip .4cm

\end{center}
\bigskip\bigskip

\begin{center} 
\textbf{Abstract}

\end{center} 
\begin{quote}  
Kaluza-Klein reductions of 11-dimensional supergravity lead to exceptional global symmetries in lower dimensions. Certain non-geometric elements of these symmetries, parameterized by a tri-vector $\gamma$, are not inherited from the higher-dimensional local symmetries, but represent  instead a symmetry enhancement produced by the isometries of the background. Here, we demonstrate how to realize this enhancement in 11 dimensions, as a symmetry principle with constrained parameters.    We show that $\gamma$ transformations exchange the equations of motion of the metric and the three-form with their Bianchi identities, in a closed form, structuring them into tri-vector multiplets. Implementing this principle as an off-shell symmetry of the theory requires the introduction of a hierarchy of dual fields, including  a six-form and a dual graviton in the initial levels.

\end{quote} 
\vfill
\setcounter{footnote}{0}
\end{titlepage}

\tableofcontents

\section{Introduction}

Kaluza-Klein (KK) reductions of 11-dimensional supergravity on $d$-dimensional tori are invariant under continuous exceptional $E_{d(d)}$ global symmetries when only the zero KK modes are kept. The $E_{d(d)}$  groups contain geometric and non-geometric elements. While the former derive from 11-dimensional diffeomorphisms and three-form gauge transformations, the latter are not associated with a symmetry of the higher-dimensional theory, but constitute instead a symmetry enhancement arising from the toroidal truncation.

Two important aspects of these exceptional groups are crucial for our work. They contain: $a$)  an $O(\tilde d, \tilde d)$ subgroup, with $\tilde d = d - 1$,  and $b$)  a tri-vector generator $\gamma$ that produces non-geometric transformations of  the descendants of the metric and the three-form, mixing them  into each other. The bi-vector components of $\gamma$, usually named $\beta$, are the generators of the non-geometric sector of $O(\tilde d, \tilde d)$. 

It has recently been demonstrated that $\beta$ acts covariantly in 10-dimensional supergravity, on the NSNS sector \cite{Beta,beta2}, in the democratic formulation of the RR sector \cite{TypeII}, and including the gauge and fermion fields of the heterotic theory \cite{Heterotic}. In essence, $\beta$ possesses a $GL(10)$ embedding that preserves the structure of the higher-dimensional fields, transforming them non-linearly. This allows to examine the invariance of the higher-dimensional action by applying $\beta$ transformations. Strictly speaking, $\beta$ does not qualify as a symmetry, as it demands the constraint $\beta^{\mu\nu} \partial_\nu = 0$, which expresses the existence of isometries. However, for practical purposes, it can be considered a symmetry principle in 10-dimensional supergravity, whose effects in $10-\tilde d$ dimensions ensure the enhancement of the internal global symmetries into the full $O(\tilde d, \tilde d)$ group.

In this paper, we extend this idea to investigate the role of the tri-vector $\gamma^{MNP}$, with $M,N=0,\dots, 10$,     
in 11-dimensional supergravity. 
Our results are presented  as follows: 
\begin{itemize}
    \item In {\bf Section 2}, we review the role of $\beta$ symmetry in the democratic formulation of 10-dimensional Type II supergravity, and explore its impact in the standard formulation. We show that this symmetry cannot be defined for the standard action, as it is not possible to assign $\beta$ transformations to the $p$-form potential fields in the standard framework. Instead, one can define the variations of the field strengths, and demonstrate that  the  equations of motion and the Bianchi identities transform into each other. This implies that $\beta$ transformations constitute an on-shell symmetry, which maps solutions into new solutions, and can thus be employed as a 
    solution-generating technique. A $\beta$ invariant action requires the introduction of dual fields within a democratic framework.

\item In {\bf Section 3}, we show that there is a unique uplift of the $\beta$ transformations of 10-dimensional Type IIA supergravity to tri-vector  transformations in 11-dimensional supergravity. We present the $\gamma$ transformations of the metric and the curvature of the three-form. As expected, we find that it is not possible to define $\gamma$ symmetry for the standard  action of 11-dimensional supergravity, but the equations of motion and the Bianchi identities transform into each other, provided the isometry constraint $\gamma^{M N P} \partial_P = 0$ is imposed. 

\item In {\bf Section 4},  we extend our results showing that the $\gamma$ transformation of the three-form (rather than its curvature)  requires the inclusion of a six-form and the extension of the global symmetries, incorporating a six-vector transformation. These correspond to the field content and  symmetries of the first non-trivial level of the  $E_{11}$ construction \cite{West}-\cite{west5}. Thus, our results  fit into this context as a minimalist bottom-up construction, with a permanent direct contact with  standard 11-dimensional supergravity.

\item In {\bf Section 5} we explore the possibility of defining the $\gamma$ transformation of the six-form through the introduction of extra degrees of freedom related to the gravitational sector. We note that the inclusion of a mixed Young tableaux field, usually interpreted as the dual graviton, leads to a natural expression for the dual of the spin connection as well as for the $\gamma$ transformation of the six-form potential.
\end{itemize}

Finally, we present some conclusions  in {\bf Section 6}. 
Details on notation, definitions and side computations are provided in the Appendices.

\section{Bi-vector $\beta$ symmetry in Type II supergravity}

In this section we discuss the role of $\beta$ symmetry in the standard formulation of 10-dimensional Type II supergravity. To this end, we first recall the democratic and standard formulations of the theory, focusing on the duality relations that connect them. Then, we review the action of $\beta$ symmetry  in the democratic formulation \cite{TypeII}, where it is well understood, with the purpose of extracting its implications for the standard formulation. Finally,  although we find that the action of the standard framework is not $\beta$ invariant, we show that the equations of motion and the Bianchi identities (BI) can be organized into $\beta$ multiplets.

\subsection{Type II supergravity}

The bosonic field content of 10-dimensional  Type II supergravity  
splits into the Neveu Schwarz-Neveu Schwarz (NSNS)  and the Ramond-Ramond (RR)  sectors. The former contains the vielbein $e_{\mu}{}^a$, the Kalb-Ramond field $b_{\mu\nu}$ and the dilaton $\phi$, and the action is given by
\begin{equation}
 S_{\rm{NSNS}} = \frac1{2\kappa_{10}}\int d^{10}x\;\sqrt{-g} e^{-2 \phi}  \left( R + 4 \Box \phi - 4 (\nabla\phi)^2-\frac{1}{12} H_{\mu\nu\rho} H^{\mu \nu \rho}
\right)  \, ,  \label{SNS}
\end{equation}
where $
H_{\mu\nu\rho}\ =\ 3\partial_{[\mu}b_{\nu\rho]}$.

The RR sector includes $p$-form potentials $D_{\mu_1\cdots\mu_p}^{(p)}$, where the allowed values of $p$ depend on the theory and on the formulation. To fix our conventions, in the following subsections we review the standard and the universal democratic formulations of the RR sector of Type IIA and Type IIB supergravities.

\subsubsection{Standard formulation }
In the standard formulation of 10-dimensional Type II supergravity, the $p$-form potentials $D_{\mu_1\cdots\mu_p}^{(p)}$ contain $p=1, 3$ in Type IIA  and  $p=0,2,4$ in Type IIB, 
respectively ruled by the actions
\begin{eqnarray}
S_{\rm{RR}}^{\rm{IIA}} 
&=&-\frac1{4\kappa_{10}}  \int d^{10}x\;\sqrt{-g}  \left( |F^{(2)}|^2 + |F^{(4)}|^2
\right)\nn\\&&+ \frac1{4\kappa_{10}} \int b\wedge d\left(D^{(3)}-b\wedge D^{(1)}\right)\wedge d\left(D^{(3)}-b\wedge D^{(1)}\right) \ , \label{RRIIA}
\end{eqnarray}
and 
\begin{eqnarray}
S_{\rm{RR}}^{\rm{IIB}} &=& -\frac1{4\kappa_{10}}\int d^{10}x \sqrt{-g}(|F^{(1)}|^2 + |F^{(3)}|^2 + \frac{1}{2}|F^{(5)}|^2) \nn\\
&&+ \frac{1}{4\kappa_{10}} \int b\wedge d\left(D^{(4)}-\frac12 b\wedge D^{(2)}\right)\wedge d(D^{(2)}-b\wedge D^{(0)}) \ , \label{RRIIB}
\end{eqnarray}
 where
\begin{eqnarray}
|F^{(n)}|^2 \ =\ \frac{1}{n!} F^{(n)\mu_1 \dots \mu_n} F^{(n)}_{\mu_1 \dots \mu_n}\, \ \ {\rm with } \ \ F^{(n)}\ = \ \ \frac1{n!}F^{(n)}_{\mu_1\cdots\mu_n}dx^{\mu_1}\wedge\cdots\wedge dx^{\mu_n}\, ,
\end{eqnarray}
and the $p = 4$ field is constrained to satisfy the  self-duality condition $F^{(5)}=\star  F^{(5)}$. The Hodge star is defined as \begin{equation}\star  F^{(n)}=\frac1{(10-n)!n!}\varepsilon_{\nu_1\cdots\nu_n\mu_1\cdots\mu_{10-n}}F^{(n)\nu_1\cdots\nu_n}dx^{\mu_1}\wedge\cdots\wedge dx^{\mu_{10-n}}\ ,\end{equation} and the field strengths can be written  as the formal sum
\be
F=e^{-b}\wedge dD\, ,
\ee
so that in Type IIA one has 
\begin{equation}
F^{(2)}\ =\  dD^{(1)}\, ,\quad
F^{(4)}\ =\ \ dD^{(3)}-b\wedge F^{(2)}\, ,
\end{equation}
and in Type IIB 
\begin{equation}
F^{(1)}\ =\  dD^{(0)}\, ,\quad
F^{(3)}\ =\ \ dD^{(2)}-b\wedge F^{(1)}\, ,\quad
F^{(5)}\ =\ dD^{(5)}-b\wedge F^{(3)} -\frac12b\wedge b\wedge F^{(1)}\, .
\end{equation}

The equations of motion for Type IIA, calculated from the action $S_{\rm{IIA}}=S_{\rm{NSNS}}+S_{\rm{RR}}^{\rm{IIA}} $, in flat notation are:
\begin{subequations}\label{eomstandard}
\begin{align} 
{\cal C} =& \ -2 e^{-2 d}  \left( R + 4 \Box \phi - 4 (\nabla\phi)^2-\frac{1}{12} H^2
\right) =0\;, \\
\widehat {\cal E}_{ab} =& \ e^{-2d} {\cal E}_{ab} +  \Delta \mathcal{E}^{\rm{IIA}}_{ab} =0\, , \label{EA}\\
\widehat {\cal B}_{ab} =& \  e^{-2d} {\cal B}_{a b} + \Delta \mathcal{B}^{\rm{IIA}}_{ab}=0\, , \label{BA} \\
\widehat{\cal D}_{a} =& \ \sqrt[4]{-g} \left[\nabla^{b}F^{(2)}_{ba} - \frac{1}{3!} F^{(4)}_{abcd} H^{bcd}\right]=0\, , \label{D1}\\
\widehat{\cal D}_{abc} =& \  - \frac{1}{3!} \sqrt[4]{-g} \left[\nabla^{d}F^{(4)}_{abcd} - \frac{1}{3!} (\star F^{(4)})_{abcdef}H^{def}\right]=0\, , \label{D3}
\end{align}
\end{subequations}
where the unconventional measure $\sqrt[4]{-g}$  is introduced for later convenience,  the generalized dilaton $d$ is defined through $e^{-2 d}=\sqrt{-g}e^{-2 \phi}$, and 
\begin{subequations}
\begin{align} 
 \mathcal{E}_{ab} =&\ -2\left(R_{ab} + 2\nabla_a\nabla_b\phi - \frac14 H_{acd}H_b{}^{cd}\right) \, ,\label{Eab}\\
    \mathcal{B}_{ab} =&\ \frac12 \nabla_c {H^{c}}_{ab} - \nabla_c \phi {H^c}_{ab} \, ,\label{Bab}\\
    \Delta \mathcal{E}^{\rm{IIA}}_{ab} =& \ -\frac{\sqrt{-g}}{2}\left[ \left( g_{ab} |F^{(2)}|^2 - 2 F^{(2)}_{ac}{F^{(2)}}_{b}{}^{c}  \right) + g_{ab}|F^{(4)}|^2 - \frac13 F^{(4)}_{acde}{F^{(4)}}_{b}{}^{cde} \right] \, ,\\
    \Delta\mathcal{B}^{\rm{IIA}}_{ab} =& \ \frac{\sqrt{-g}}{4} \left[ F_{ab}^{(4)cd}F^{(2)}_{cd}  + \frac{1}{4!} \star  F_{abcdef}^{(4)}F^{(4)cdef} \right]    \ .
\end{align}
\end{subequations}
For Type IIB, the field equations obtained from the action $S_{\rm{IIB}}=S_{\rm{NSNS}}+S_{\rm{RR}}^{\rm{IIB}}$  are:
\begin{subequations}\label{eomIIB}
\begin{align}
\widehat{\mathcal{E}}_{ab} =&\ e^{-2d}\mathcal{E}_{ab} + \Delta \mathcal{E}^{\rm{IIB}}_{ab} =0\, , \label{EB}\\
\widehat{\mathcal{B}}_{ab} =&\ e^{-2d}\mathcal{B}_{ab} + \Delta \mathcal{B}^{\rm{IIB}}_{ab}=0\, , \label{BB}\\
    \widehat{\mathcal{D}} =& \ \sqrt[4]{-g} \left[ \nabla^a F^{(1)}_a - \frac{1}{3!}F^{(3)}_{abc}H^{abc} \right] = 0\, , \label{D0}\\
    \widehat{\mathcal{D}}_{ab} =& \ \sqrt[4]{-g}\left[  \nabla^cF_{abc}^{(3)} - \frac{1}{3!}F^{(5)}_{abcde}H^{cde} \right] = 0\, ,\label{D2}\\
 \widehat {\cal{D}}_{abcd} =& \ \frac{\sqrt[4]{-g}}{2\cdot 4!} 
 \left[\nabla^e F^{(5)}_{abcde} + \frac{1}{3!}(\star  F^{(3)})_{abcde_1e_2 e_3}H^{e_1 e_2 e_3}\right] = 0\, , \label{D4}     
\end{align}
\end{subequations}
where 
\begin{subequations}
\begin{align}
\Delta \mathcal{E}^{\rm{IIB}}_{ab} =&\ -\frac{\sqrt{-g}}{2}\left[ \vphantom{\frac 1 2}\left( g_{ab} |F^{(1)}|^2 - 2 F^{(1)}_{a}F^{(1)}_{b} \right) + g_{ab}|F^{(3)}|^2 -  F^{(3)}_{acd}{F^{(3)}}_{b}{}^{cd} \right.\nn\\
&\quad \quad \quad\quad\quad+\left. \frac{1}{2} g_{ab}|F^{(5)}|^2 - \frac{1}{24}F^{(5)}_{acdef}F^{(5)}_{b}{}^{cdef}\right]\, ,\\
 \Delta\mathcal{B}^{\rm{IIB}}_{ab} =&\ \frac{\sqrt{-g}}{2} \left[ F^{(3)}_{abc}F^{(1)c} + \frac{1}{3!} F_{ab}^{(5)cde}F^{(3)}_{cde}\right]\, ,
 \end{align}
\end{subequations}
and the self-duality relation  must be imposed after deriving the equations of motion. 

The  RR fields verify the Bianchi  identities (BI) 
\begin{eqnarray}
d\left(e^b\wedge F\right) =0     \, ,
\end{eqnarray}
which  are specifically, for Type IIA,
\be
dF^{(2)} = 0 \ , \ \ \ dF^{(4)} + H\wedge F^{(2)} = 0\ , \label{BI2A}
\ee
and for Type IIB,
\be
dF^{(1)} = 0 \ , \ \ \ dF^{(3)} +H\wedge F^{(1)} = 0 \ , \ \ \ dF^{(5)} + H \wedge F^{(3)} = 0\ . \label{BI2B}
\ee

\subsubsection{Democratic formulation}
In the universal democratic formulation of Type II supergravity,  the “electric” and “magnetic” potentials of all RR fields are treated on equal footing. 
The $p$-form potentials $D_{\mu_1\cdots\mu_p}^{(p)}$ include $p=1, 3, 5, 7$    in Type IIA and $p=0,2,4,6,8$ in Type IIB. These fields obey the pseudo-action \cite{bko}
\begin{equation}
    S_{\rm{II}} = S_{\rm{NSNS}} + S_{\rm{RR}}^{\rm{II}} \ ,
\end{equation}
where
\begin{equation}
S_{\rm{RR}}^{\rm{II}} =-\frac1{8\kappa_{10}}\int d^{10}x\sqrt{-g}\sum_n|F^{(n)}|^2= \frac1{8\kappa_{10}} \int \sum_{n} F^{(n)}\wedge \star  F^{(n)} \  ,
\end{equation}
with
$n=2,4,6,8$  in type IIA and $n=1,3,5,7,9$ in Type IIB.

The equations of motion are obtained by treating all the RR potentials as independent fields. In  flat notation they take the form  
\begin{subequations} \label{eomdemocratic}
\begin{align} 
{\cal C} =& \ -2 e^{-2 d}  \left( R + 4 \Box \phi - 4 (\nabla\phi)^2-\frac{1}{12} H^2
\right) =0\;, \label{C}\\
\widetilde {\cal E}_{ab} =&  \  e^{-2d} {\cal E}_{ab} 
-\frac14  \sqrt{-g} \sum_{n}\frac{1}{n!} \left(g_{ab} F^{(n)}_{c_1 \dots c_n} F^{(n)}{}^{c_1 \dots c_n} - 2 n F^{(n)}_{a c_1 \dots c_{n-1}} F^{(n)}_{b}{}^{ c_1 \dots c_{n-1}} \right)\, =0\;,   \label{E}
 \\ \widetilde {\cal B}_{ab} =& \ e^{-2d} {\cal B}_{a b} + \frac14 \sqrt{-g} \sum_{n} \frac{1}{(n-2)!}F^{(n)}_{a b c_1 \dots c_{n-2}} F^{(n)}{}^{c_1 \dots c_{n-2}}\, =0 \;, \label{B}\\
\widetilde {\cal E}^{(p)}_{a_1 \dots a_p }=& \ -\frac{\sqrt{-g}}{2\cdot p!} \left[\star d\left(e^{-b}\wedge \star F\right)\right]_{a_1 \dots a_p}  = 0 \;, \label{CalE}
\end{align}
\end{subequations}
where ${\cal E}_{ab}$ and ${\cal B}_{ab}$ coincide with those defined in the standard formulation (\ref{Eab}) and (\ref{Bab}), respectively.

The logic of this formulation comes from  supplementing the field equations with the duality relations of the RR curvatures
\begin{eqnarray}
  \star F^{(n)} = (-)^{n(n-1)/2}F^{(10-n)}\, . \label{dualityrel}
\end{eqnarray}

In Type IIA, one must replace the curvatures $F^{(8)}$ and $F^{(6)}$  in  (\ref{eomdemocratic}) by $F^{(2)}$ and $F^{(4)}$, respectively,  using (\ref{dualityrel}). After dualization, 
 (\ref{C}) remains invariant, (\ref{E}) and (\ref{B}) become (\ref{EA}) and (\ref{BA}), respectively, $\widetilde {\cal E}^{(1)}$ and $\widetilde {\cal E}^{(3)}$ in  (\ref{CalE}) become (\ref{D1}) and (\ref{D3}), respectively, and finally, $\star \widetilde {\cal E}^{(5)}$ and $\star \widetilde {\cal E}^{(7)}$  in (\ref{CalE}) become the BI (\ref{BI2A}), which in flat indices read 
\begin{subequations}\label{BIs}
\begin{align} 
 {\cal I}_{abcde} =& \  \sqrt[4]{-g}\left(5\nabla_{[a}F^{(4)}_{bcde]}+ \frac{5!}{2! \cdot 3!} H_{[abc}F^{(2)}_{de]} \right) =0\, ,\\
 {\cal I}_{abc} =& \ \sqrt[4]{-g} \; 3\;  \nabla_{[a}F^{(2)}_{bc]}=0\ .
\end{align}
\end{subequations}

In Type IIB, one must replace the curvatures $F^{(9)}$ and $F^{(7)}$  in  (\ref{eomdemocratic}) by $F^{(1)}$ and $F^{(3)}$ using (\ref{dualityrel}), and self-dualize $F^{(5)}$. After dualization, (\ref{C}) remains invariant, and (\ref{E}) and (\ref{B}) become (\ref{EB}) and (\ref{BB}), respectively. The field equations $\widetilde {\cal E}^{(0)}$, $\widetilde {\cal E}^{(2)}$ and $\widetilde {\cal E}^{(4)}$ in  (\ref{CalE}) become (\ref{D0}), (\ref{D2}) and (\ref{D4}), respectively. Finally, $\star \widetilde {\cal E}^{(4)}$, $\star \widetilde {\cal E}^{(6)}$ and $\star \widetilde {\cal E}^{(8)}$  in (\ref{CalE}) become the BI (\ref{BI2B}), which in flat indices read 
\begin{subequations}\label{biIIB}
\begin{align}
 {\cal I}_{abcdef} =&\ \sqrt[4]{-g}\left( 6\nabla_{[a}{F}^{(5)}_{bcdef]} + \frac{6!}{3! \; 3!} H_{[abc}F^{(3)}_{def]}\right)\, ,\\
{\cal I}_{abcd} =&\ 4 \sqrt[4]{-g}\left( \nabla_{[a}F^{(3)}_{bcd]} +  H_{[abc}F^{(1)}_{d]}\right)  \, ,\\
 {\cal I}_{ab} =&\ 2 \sqrt[4]{-g} \; \nabla_{[a}F^{(1)}_{b]} \, .
\end{align}
\end{subequations}

\subsection{$\beta$ symmetry in Type II supergravity}

The purpose of this section is to understand the action of  $\beta$ symmetry  in the standard formulation of Type II supergravity. To this end, we first review how it works in the democratic formulation \cite{TypeII}, and then use the results of the previous subsections to analyze its behavior in the standard version.

\subsubsection{Democratic formulation}
The $\beta$ transformations of the fields in the democratic formulation of Type II theories are  \cite{TypeII}  
\begin{subequations}\label{betademoc}
\begin{align}
\delta_\beta e_\mu{}^a &=\ -e_\mu{}^bb_{bc}\beta^{ca}\, ,\\
\delta_\beta b_{\mu \nu} &=\ -\beta_{\mu\nu}-b_{\mu\rho}\beta^{\rho\sigma}b_{\sigma\nu}\, ,\\
\delta_\beta \phi &=\ \frac12\beta^{\mu\nu}b_{\mu\nu}\, , \\
\delta_\beta D^{(p)}_{\mu_1\cdots\mu_p} &= -\frac12\beta^{\rho\sigma}D^{(p+2)}_{\rho\sigma\mu_1\cdots\mu_p}\, . \label{betatransfIIAfields}
\end{align}
\end{subequations}
 These transformations must be supplemented with an isometry constraint
\begin{equation}
\beta^{\mu \nu} \partial_\nu \dots = 0 \ ,\label{isometrybeta}
\end{equation}
where the dots refer to any field and/or gauge parameter in the theory. The transformations \eqref{betademoc} imply
\begin{subequations}\label{betaIIAdemoc}
\begin{align}
\delta_\beta \left( \sqrt{- g} e^{- 2 \phi} \right)\ &= \ 0 \, , \\
\delta_\beta \omega_{cab}\ &=\ \beta_{[a}{}^dH_{b]cd}-\frac12\beta_c{}^dH_{abd}  \, ,\\
\delta_\beta H_{abc}\ &=\ - 3 \nabla_{[a}\beta_{b c]}  \, ,\\
\delta_\beta(\nabla_a\phi)\ &=\ \frac12\beta^{cd}H_{acd}  \, ,\\
\delta_\beta( \nabla_a \nabla_b\phi) \ &=\ \frac12\nabla_{(a}\left(\beta^{cd} H_{b)cd}\right) - \beta^{c}{}_{(a} H_{b) c d} \nabla^{d}\phi  \, , \\
\delta_\beta R \ &=\ - 2\nabla^a\left( \beta^{b c}  H_{a b c}\right)-\frac12 \nabla^{a} \beta^{b c} H_{a b c}\, , \\
\delta_{\beta} F^{(n)}_{a_1\cdots a_n} &=
-\frac12 \beta^{cd} b_{cd} F^{(n)}_{a_1\cdots a_n}+\frac{n(n-1)}2\beta_{[a_1a_2}F^{(n-2)}_{a_3\cdots a_n]}-\frac12\beta^{cd}F^{(n+2)}_{cda_a\cdots a_n}\, . \label{transff}
\end{align}
\end{subequations}
These $\beta$ variations mix the equations of motion and the BI as follows \cite{TypeII}, 
\begin{subequations}\label{transfIIA}
\begin{align}
\delta_{\beta} \widetilde{\mathcal{E}}_{ab} =& \ -4\beta_{c(a}\widetilde{\mathcal{B}}_{b)}{}^{c} \;,
\;\;\;\;\;\;\;\;\;\;\;\;\;\;\;\;\;\;\;\;\;\;\;
\delta_\beta \widetilde{\mathcal{B}}_{ab} = \beta_{c[a}\widetilde{\mathcal{E}}_{b]}{}^{c}\, ,\\
\delta_\beta \widetilde{\cal E}^{(p)}_{a_1 \dots a_p} =&  \frac{p(p-1)}{2} \beta_{[a_1 a_2}\; \widetilde{\cal E}^{(p-2)}_{a_3 \dots a_p]}  - p \; \beta_{[a_1}{}^{c} \widetilde{\cal E}^{(p)}{}^{d}{}_{a_2\dots a_p]} \; b_{cd}\;,  
\end{align}
\end{subequations}
where it is implicitly assumed that $\widetilde{\cal E}^{(p-2)}$ vanishes for $p<2$. $\delta_{\beta}{\cal C}$ is not included here because ${\cal C}\sim S_{\rm{NSNS}}$ is $\beta$ invariant. 

We conclude the discussion on $\beta$ symmetry in the democratic formulation of Type II supergravity with a comment on the consistency between duality relations and $\beta$ transformations. The combination 
\begin{eqnarray}
\Gamma^{(n)}= F^{(n)} + (-)^{\frac{n(n+1)}{2} } \star  F^{(10-n)}    \, , \quad 0\leq n\leq9\, ,
\end{eqnarray}
which vanishes as a consequence of the duality relations  (\ref{dualityrel}), transforms as
\begin{eqnarray}
\delta_\beta \Gamma^{(n)}_{\mu_1 \dots \mu_n} &=& 
-\frac12 \beta^{\rho \sigma} b_{\rho \sigma} \Gamma^{(n)}_{\mu_1 \dots \mu_n}- n\  b_{[\mu_1 |\rho|} \beta^{\rho\sigma} \Gamma^{(n)}_{|\sigma|\mu_2 \dots \mu_n]} \cr && 
+ \frac{n(n-1)}{2} \beta_{[\mu_1 \mu_2} \Gamma^{(n-2)}_{\mu_3 \dots \mu_n]} -\frac12 \beta^{\rho \sigma} \Gamma^{(n+2)}_{\rho \sigma \mu_1 \dots \mu_n} \, .\label{deltaGamma}
\end{eqnarray}
Hence, $\beta$ transformations preserve the duality condition $\Gamma^{(n)}=0$. This property will be important when  trying to apply this symmetry principle in the standard formulation of the theory.


\subsubsection{Standard formulation}
Due to the $\beta$ covariance  of the duality relations, the variations (\ref{betaIIAdemoc})  also apply to the equations of motion (\ref{eomstandard}), \eqref{eomIIB} and the BI (\ref{BIs}), \eqref{biIIB} in the standard formulation.
In particular, applying the duality relations to (\ref{transff}) yields the variations of the RR curvatures of Type IIA
\begin{eqnarray}
\delta_\beta F^{(2)}_{ab} &=& 
-\frac12 \beta^{cd} b_{cd} F^{(2)}_{ab} 
- \frac12 \beta^{cd} F^{(4)}_{cdab}\,  , \\
\delta_\beta F^{(4)}_{a_1 a_2 a_3 a_4} &=& 
-\frac12 \beta^{cd} b_{cd} F^{(4)}_{a_1 a_2 a_3 a_4}  
+ 6 \beta_{[a_1 a_2} F^{(2)}_{a_3 a_4]}
- \frac12 \beta^{cd} \left(\star  F^{(4)} \right)_{cd a_1 a_2 a_3 a_4}\, ,
\label{betaIIAstand}
\end{eqnarray}
and Type IIB
\begin{eqnarray}
\delta_\beta F^{(1)}_{a} &=& 
-\frac12 \beta^{cd} b_{cd} F^{(1)}_{a} 
- \frac12 \beta^{cd} F^{(3)}_{cda}\, , \\ 
\delta_\beta F^{(3)}_{a_1 a_2 a_3} &=& 
-\frac12 \beta^{cd} b_{cd} F^{(3)}_{a_1a_2a_3}  
+ 3 \beta_{[a_1 a_2} F^{(1)}_{a_3]} - \frac12 \beta^{cd} F^{(5)}_{cda_1a_2 a_3}\, ,\\
\delta_\beta F^{(5)}_{a_1 ... a_5} &=& 
-\frac12 \beta^{cd} b_{cd} F^{(5)}_{a_1 ... a_5}  
+ 10 \beta_{[a_1 a_2} F^{(3)}_{a_3 ... a_5]} + \frac12 \beta^{cd} \left(\star  F^{(3)}\right)_{cda_1 ... a_5}\, .
\end{eqnarray}

Two important results on the action of $\beta$ symmetry  in the standard formulation can be obtained:
\begin{enumerate}
\item  The field equations (\ref{eomstandard}),  \eqref{eomIIB} and the BI (\ref{BIs}), \eqref{biIIB} form a closed set of equations under  $\beta$ transformations. Explicitly, for Type IIA one has
\begin{subequations}
\begin{align}
\delta_{\beta} \widehat{\mathcal{E}}_{ab} =&\ -4\beta_{c(a} \widehat{\mathcal{B}}_{b)}{}^{c} \;,
\;\;\;\;\;\;\;\;\;\;\;\;\;\;\;\;\;\;\;\;\;
\delta_\beta \widehat{\mathcal{B}}_{ab} = \beta_{c[a} \widehat{\mathcal{E}}_{b]}{}^{c}\, ,\\
\delta_{\beta} \widehat{\mathcal{D}}_{a} =&\ 
3 \beta^{bc} \widehat{\cal D}_{abc} \;,
\;\;\;\;\;\;\;\;\;\;\;\;\;\;\;\;\;\;\;\;\;\;\;\;
\delta_{\beta} \widehat{\cal D}_{abc} = - \frac12 \beta_{[ab}\widehat{\cal D}_{c]}
+ \frac{1}{12} \beta^{de} (\star \widehat{\cal I}_5)_{abcde}\, ,\\
\delta_{\beta} \widehat{\cal I}_{abc} = &\ -\frac12 \beta^{de} \; \widehat{\cal I}_{abcde}
\;,
\;\;\;\;\;\;\;\;\;\;\;\;\;\;\;
\delta_{\beta} \widehat {\cal I}_{abcbd} =  10 \beta_{[ab}\widehat {\cal I}_{cde]} 
- 3 \beta^{fg} (\star \widehat{\cal D}_{3})_{abcdefg}
\;\;\, , \label{TrStIIA}
\end{align}\label{TrStIIAb}
\end{subequations}
and for Type IIB
\begin{subequations}
\begin{align}
\delta_{\beta} \widehat{\mathcal{E}}_{ab} =&\ -4\beta_{c(a} \widehat{\mathcal{B}}_{b)}{}^{c}
\;,\;\;\;\;\;\;\;\;\;\;\;\;\;\;\;\;\;\;\;
\delta_\beta \widehat{\mathcal{B}}_{ab} =\beta_{c[a} \widehat{\mathcal{E}}_{b]}{}^{c} \, , \\
\delta_\beta {\widehat {\cal D}} =& \ - \frac12 \beta^{ab} \widehat {\cal D}_{ab} \;,\;\;\;\;\;\;\;\;\;\;\;\;\;\;\;\;\;\;\;\;
\delta_\beta {\widehat {\cal D}}_{ab} = \beta_{ab} \widehat{\cal D} - \frac12 \beta^{cd} \widehat{\cal D}_{abcd}\, ,\\
\delta_\beta {\widehat {\cal D}}_{abcd} =&\ 6 \beta_{[ab}\widehat{\cal D}_{cd]} -\frac12 \beta^{ef} (\star \widehat{\cal I}_4)_{abcdef}\, ,\\
\delta_\beta {\widehat {\cal I}}_{ab} =&\ -\frac12 \beta^{cd} \widehat {\cal I}_{abcd}\;, \;\;\;\;\;\;\;\;\;\;\;\;\;
\delta_\beta {\widehat {\cal I}}_{abcd} = 6 \beta_{[ab} \widehat {\cal I}_{cd]} + \frac12 \beta^{ef} (\star \widehat {\cal D}_{4})_{abcdef}\, . \label{TrStIIB}
\end{align}
\end{subequations}

\item Importantly, while the  BI $\widehat{\cal I}_{abc} = 0$ is  $\beta$ invariant off-shell, we find that $\widehat{\cal I}_{a_1\dots a_5} = 0$ is only on-shell  invariant in the standard formulation of Type IIA, due to the presence of the equation of motion $\widehat{\cal D}_3$ in its transformation (\ref{TrStIIA}). This implies that off-shell there is no way to assign a $\beta$ transformation to $D^{(3)}$, and then  there is no chance to asses $\beta$ symmetry in the standard action (\ref{RRIIA}), because $D^{(3)}$ appears explicitly (and not only through its curvature) in the Chern-Simons term. The same conclusion applies for the action of Type IIB \eqref{RRIIB}, since $\delta_\beta {\widehat {\cal I}}_{abcd}$ depends on $\widehat {\cal D}_{4}$. 

There are other ways to reach the same result. On the one hand,  it is not possible to extract the $\beta$ transformation for $D^{(3)}$ from (\ref{transff}).  The reason is that since $F^{(4)} = d D^{(3)} - b \wedge F^{(2)}$, it turns out that $\delta_{\beta} \left(F^{(4)} +  b \wedge F^{(2)} \right)$ is not exact, and indeed it is not even closed (off-shell).  Then, in the standard formulation of Type IIA, one is forced to work with the curvature field $F^{(4)}$ instead of its potential $D^{(3)}$. 
On the other hand, the duality relation cannot be imposed on $\delta_\beta D^{(3)}$ in (\ref{betatransfIIAfields}), because it depends on the dual potential $D^{(5)}$. Hence, a consistent $\beta$ variation for $D^{(3)}$ requires the presence of dual potentials. 
Equivalent arguments  apply to $D^{(4)}$ in Type IIB.
\end{enumerate}

 In summary,  in the standard formulation of 10-dimensional Type II supergravity,  $\beta$ transformations can be defined for the NSNS fields and for the curvatures of the $p$-form potentials. The equations of motion and the BI are transformed into each other, in a closed form.  Defining $\beta$ transformations for the $p$-form potentials requires a democratic formulation with dual fields. In the following sections we will explore the uplift of this symmetry to 11-dimensional supergravity.

\section{Tri-vector $\gamma$ symmetry in 11-dimensional supergravity}\label{TrivectorSymmetry}

The 
bosonic field content of the standard formulation of 11-dimensional supergravity comprises the metric  $G_{M N}$ and a three-form $A_{M N P}$, and the action is given by
\begin{eqnarray}
 S_{11} &=&\frac1{2\kappa_{11}^2}\int d^{11}x\;  \sqrt{-G} \left(\vphantom{\frac12}\right. R-\frac{1}{2\cdot 4!} F_{MNPQ} F^{MNPQ} \cr
&& \ \ \ \ \ \ \ \ \  \ \ \ \ \ \ \ \ \ \ \   - \left(\frac{2}{4!}\right)^4 \varepsilon^{M_1 ... M_{11}}A_{M_1 M_2 M_3} F_{M_4 M_5 M_6 M_7} F_{M_8 M_9 M_{10} M_{11}}\left.\vphantom{\frac12}\right)\ , \label{s11}
\end{eqnarray}
with  $ F_{M N P Q} = 4\partial_{[M}  A_{M N P]}$.
 A circle reduction of this theory gives rise to the standard formulation of 10-dimensional Type IIA supergravity,  with the following KK decomposition
\begin{subequations}
\begin{align}
G_{\mu  \nu} =&\ e^{-\frac23 \phi} g_{\mu\nu} + e^{\frac43 \phi} D^{(1)}_{\mu} D^{(1)}_{\nu} \, ,\\
G_{\mu  10} =&\ - e^{\frac43 \phi} D^{(1)}_{\mu}\, , \\
G_{10 10} =&\ e^{\frac43 \phi}\, , \\
A_{\mu \nu \rho} =&\ - \left(D^{(3)}_{\mu \nu \rho} - 3\; b_{[\mu \nu} D^{(1)}_{\rho]}\right)\, ,\\ 
A_{\mu \nu 10} =&\ - b_{\mu \nu} \, .
\end{align}\label{ansatz}
\end{subequations}

 The uplift of the 10-dimensional bi-vector $\beta^{\mu \nu}$  will be a trivector $\gamma^{M N P}$ in 11 dimensions,  with
\begin{equation}
\gamma^{\mu \nu 10} = - \beta^{\mu \nu}\ .
\end{equation}
$\gamma$ is a generator of the non-geometric sector of $E_{d(d)}$, where non-geometric refers to the fact that it mixes the gravitational and $p$-form scalar fields.
In addition, it  contains a 10-dimensional tri-vector component $\gamma^{\mu \nu\rho}$, whose effect we ignored in the previous section. The  isometry constraint (\ref{isometrybeta}) uplifts to a  $GL(11)$ covariant constraint
\begin{equation}
\gamma^{M N P} \partial_P\dots = 0 \ ,
\end{equation}
where again, the dots represent any field or gauge parameter. 

In this section we explore the action of $\gamma^{M NP}$ transformations on the bosonic fields of the standard formulation of 11-dimensional supergravity, building on  the results of the previous section, where we analyzed in detail  the action of $\beta$ transformations  on the right hand side of (\ref{ansatz}). We start with the gravitational sector, recalling  the  $\beta$ variations  of the 10-dimensional components of $G_{M N}$ (\ref{betademoc}) 
\begin{subequations}
\begin{align}
\delta_\beta \phi =&\ \frac12 \beta^{\mu\nu} b_{\mu\nu}\, ,\\
\delta_\beta g_{\mu\nu} =&\ -  g_{\mu \rho} \beta^{\rho \sigma} b_{\sigma \nu}
-  g_{\nu \rho} \beta^{\rho \sigma} b_{\sigma \mu}\, , \\
\delta_\beta D^{(1)}_{\mu} =&\ -\frac12 \beta^{\rho\sigma} D^{(3)}_{\mu \rho \sigma} \ .
\end{align}
\end{subequations}
It is easy to see that the only possible uplift of these transformations to 11 dimensions is given by
\begin{eqnarray}
\delta_\gamma G_{MN}= G_{R(M} \gamma^{RPQ} A_{N)PQ} - \frac19 G_{MN} \gamma^{PQR} A_{PQR}\ ,  
\end{eqnarray}
after taking  $\gamma^{\mu\nu\rho}=0$. This determines the $\gamma$ transformation of the 11-dimensional vielbein,  
\begin{equation}
    \delta_\gamma E_{M}{}^{A} = -\frac{E_M{}^A}{18} A_{NPQ}\gamma^{NPQ} + \frac12 E_M{}^B \gamma^{ADE}A_{BDE}\ , \label{gammaE}
\end{equation}
which is defined up to  Lorentz transformations.

We cannot follow the same procedure to find the $\gamma$ transformation of the 3-form $ A_{M N P}$, because $\delta_\beta D^{(3)}$ is not available in the standard formulation of Type IIA supergravity, as discussed at the end of the previous section. However, we can try to uplift  the transformation of its curvature $\delta_\beta F^{(4)}$, expecting that the circle reduction of the $\gamma$ variation of $ F_{M N P Q} = 4\partial_{[M}  A_{M N P]}$ reproduced the $\beta$ transformations of $H_{\mu\nu\rho}$  and $F^{(4)}_{\mu\nu\rho\sigma}$ in (\ref{betaIIAdemoc}).  We find by inspection that the only possibility is given by
\begin{eqnarray}
\delta_\gamma  F_{MNPQ} &=&  - \frac{1}{3!}  A_{RST} \gamma^{RST} {F}_{MNPQ} 
- 2 \gamma^{RST}  A_{RS[M} {F}_{NPQ]T} 
\cr
&& -  4\; \partial_{[M}\gamma_{NPQ]} 
+ \frac{1}{3!} \gamma^{RST} (\star F)_{RSTMNPQ} \;.   \label{deltaF11}
\end{eqnarray}
Indeed, considering that $ F_{10\mu\nu\rho}= H_{\mu\nu\rho}$ and $F_{\mu\nu\rho\sigma} = -F^{(4)}_{\mu\nu\rho\sigma} + \left(H\wedge D^{(1)}\right){}_{\mu\nu\rho\sigma}$,  one obtains the expected transformation rules of the Type IIA components
\begin{eqnarray}
\delta_\beta H_{\mu\nu\rho} &=& - (d\beta)_{\mu\nu\rho} + 3\beta^{\sigma\lambda} b_{\sigma[\mu} H_{\nu\rho]\lambda}\, , \\
\delta_\beta F^{(4)}_{\mu \nu \rho \sigma} &=& 
-\frac12 \beta^{\lambda\tau} b_{\lambda\tau} F^{(4)}_{\mu \nu \rho \sigma}  
-4\;  \beta^{\lambda\tau}  b_{\lambda[\mu}F^{(4)}_{\nu \rho \sigma]\tau} 
+ 6 \; \beta_{[\mu \nu} F^{(2)}_{\rho \sigma]}
- \frac12 \beta^{\lambda\tau} \left(\star F^{(4)}\right){}_{\lambda \tau \mu \nu \rho \sigma}  \ , \ \ \ \ 
\end{eqnarray}
when $\gamma^{\mu\nu\rho} = 0$.

Although the transformations $\delta_\gamma E_M{}^A$ and $ \delta_\gamma F_{M N P Q}$ are now available,  we are unable to explicitly explore  $\gamma $ symmetry in the  action  \eqref{s11} due to the absence of  the transformation for the three-form $\delta_\gamma A_{MNP}$.  This limitation was expected, as the Type IIA truncation lacks off-shell invariance. Additionally, we demonstrate in Appendix \ref{AppNoGo}  that the circle reduction of the most general proposal for tri-vector transformations of the metric and the three-form potential is not compatible with the $\beta$ transformation  rules of the NSNS fields of Type IIA supergravity. 

We can however hope that the Bianchi identity
\begin{eqnarray}
{\cal Z}^{(5)}_{ABCDE}= \sqrt{-G } \;(dF)_{ABCDE} = 0 \  \label{BI11}
\end{eqnarray}
and the equations of motion
\begin{subequations}\label{eom11}
\begin{align}
\mathcal{E}^{(2)}_{AB} =& \ -2 \sqrt{-G} \left(R_{AB}-\frac{1}{12} F_{(A}{}^{CDE}F_{B)CDE} +\frac{1}{6} \eta_{AB} |F|^2 \right)  = 0 \;,\label{eomeab}\\
\mathcal{E}^{(3)}_{ABC} =& \
\frac{\sqrt{-G}}{3!} \left(\nabla_{D}F^{D}{}_{ABC} -\frac{1}{48} (\star F)_{ABC D_1 D_2 D_3 D_4} F^{D_1 D_2 D_3 D_4} \right) = 0\ , \label{eomeabc} 
 \end{align}
\end{subequations}
transform into each other through $\gamma$ variations. 
Before examining this question, it is instructive to look at the $\gamma$ transformations of the tensors and connections that are involved, namely
\begin{subequations}
\begin{align}
    \delta_\gamma \omega_{ABC} =& \ \frac{(A \cdot \gamma)}{18} \omega_{ABC} - \frac{1}{9} \eta_{A[B}(F\cdot \gamma)_{C]} + \frac{1}{2}\gamma_{[B}{}^{DE}F_{C]ADE} - \frac{1}{4}\gamma_{A}{}^{DE}F_{BCDE}
\, ,\\
 \delta_\gamma R_{AB} =& \  \frac{(A\cdot \gamma)}{9} R_{AB} + \frac{\eta_{AB}}{18} \nabla^{C}(F\cdot \gamma)_C + \frac{1}{2} \nabla^{C}[\gamma^{EF}{}_{(A}F_{B)CEF}] - \frac{1}{6}\nabla_{(A}[\gamma^{CDE}]F_{B)CDE} \ , \\
     \delta_\gamma F_{ABCD} =& \  \frac{1}{18} (A\cdot \gamma) F_{ABCD} - d\gamma_{ABCD} + \frac{1}{6} \gamma^{FGH} (\star  F)_{FGH ABCD} \ ,
 \end{align}
\end{subequations}
where $A \cdot \gamma=A_{MNP}\gamma^{MNP}$ and $(F\cdot \gamma)_{M} = F_{MNPQ}\gamma^{NPQ}$.
We see that $\gamma$ transforms  the geometric sector into the gauge sector, exposing its non-geometric nature, and suggesting an underlying 11-dimensional generalized geometric structure.

Applying these variations to the BI (\ref{BI11}) and the field equations (\ref{eom11}), we find that they transform into each other, as expected:
\begin{subequations}\label{gammaeom11}
\begin{align}
     \delta_\gamma \mathcal{E}^{(2)}_{AB} =& \  -\frac{2}{3} \eta_{AB}\gamma^{FGH} \mathcal{E}^{(3)}_{FGH} + 6\, \gamma^{EF}{}_{(A} \, \mathcal{E}^{(3)}{}_{B)EF} \;, \label{gammag}\\
      \delta_\gamma \mathcal{E}^{(3)}_{ABC} =& \ -\frac{1}{2} \mathcal{E}^{(2)}_{E}{}^{[A} \gamma^{BC]E} - \frac{1}{(3!)^2}  \; \gamma^{DEF}(\star  {\cal Z}^{(5)})_{DEFABC} \ , \\
    \delta_\gamma {\cal Z}^{(5)}_{ABCDE} &= \ \gamma^{FGH}(\star  \mathcal{E}^{(3)})_{FGH ABCDE} \ .  \label{deltaZ}
 \end{align}
\end{subequations}
These equations are the $GL(11)$ covariant uplift of (\ref{TrStIIAb}).

We then reach similar  conclusions for $\gamma$ symmetry to the ones displayed at the end of the previous section for $\beta$ symmetry in the standard formulation.   The equations of motion  (\ref{eom11}) and the BI   (\ref{BI11}) constitute a closed set of equations under $\gamma$ transformations, as can be seen in (\ref{gammaeom11}). Importantly, we find that ${\cal Z}^{(5)}_{ABCDE} = 0$ is only on-shell invariant, owing to the presence of the field equation ${\cal E}^{(3)}$ in its transformation $\delta_\gamma {\cal Z}^{(5)}$  (\ref{deltaZ}). This implies that there is no chance to realize $\gamma$ transformations as an off-shell symmetry of the action (\ref{s11}), which is independently verified in Appendix \ref{AppNoGo}  without invoking RR fields. 

 It is also not possible to extract the $\gamma$ transformation for $A_{MNP}$ from (\ref{deltaF11}), because $\delta_\gamma F_{MNPQ}$ is not exact, and indeed it is not even closed, as follows from (\ref{deltaZ}). Then, $\gamma$ symmetry in the standard formulation  requires to work with the curvature field $F_{MNPQ}$, and not with the potential $A_{MNP}$.
 
Following the lessons of the previous section, we can then speculate that examining $\gamma$ invariance of the action requires a democratic formulation with dual fields.

\section{Diving into the  level decomposition of $E_{11}$}

We have seen that tri-vector $\gamma$ transformations cannot be defined for the potential $A_{MNP}$ in the standard formulation of 11-dimensional supergravity. The obstruction is equivalent to what we found for the $\beta$ transformations of $D^{(3)}$ and $D^{(4)}$ in the standard formulation of 10-dimensional Type IIA and IIB supergravity, where it was circumvented  by moving towards a democratic formulation, introducing extra fields and duality relations. In
this section we show  that the addition of  a six-form $A_6$ to the bosonic field content of 11-dimensional supergravity allows to assess the $\gamma$ transformation of the three-form $A_3$. This paves the way to study closure conditions, which further imply the existence of an 11-dimensional six-vector global symmetry.

 The importance of the six-form gauge field in 11-dimensional supergravity was first
 realized in \cite{nicolai,DAuria}, and was later related to central charges in the M-theory
 superalgebra  in \cite{townsend}. Actually, a dual formulation of the standard  theory containing a six-form $A_6$ is known to be necessary,  since the M-theory membrane couples naturally to the three-form, but the coupling of the magnetic dual five-brane requires a seven-form field strength $F_{M_1\cdots M_7}$.

To see the impact of including $A_6$ on $\gamma$ symmetry, we begin by considering the field equation for $A_{MNP}$ \eqref{eomeabc}, which can be conveniently written as 
\begin{eqnarray}
0=(\star \mathcal{E}^{(3)})_{M_1 ... M_{8}} =
-\left[d(\star F_4) + \frac12 F_4\wedge F_4\right]{}_{M_1 \dots M_8}\; .
\label{3formEOM}
\end{eqnarray}
This equation is reinterpreted as the BI of a dual curvature  $F_7=\star F_4$ 
\begin{eqnarray}
dF_7 + \frac12 F_4\wedge F_4= 0\;.
\label{BIF7}
\end{eqnarray}
Its potential $A_6$ is now considered as an independent field, defined from the  equation above as
\begin{eqnarray}
F_7=dA_6 - \frac12 A_3\wedge F_4 \; \label{F7}  
\end{eqnarray} (see \cite{West:2011mm} for an alternative derivation    motivated from $E_{11}$).

We can now try to define the $\gamma$ transformation for $A_3$. Recall that it was not possible to define $\delta_\gamma A_3$ in the standard formulation because this would imply $\delta_\gamma{\cal Z}^{(5)}=\delta_\gamma(dF_4)= d^2(\delta_\gamma A_3)=0$, while we found  that $\delta_\gamma{\cal Z}^{(5)}=0$ only holds on-shell. However,  the inclusion of $A_6$ provides extra tools to bypass this issue. Indeed,   $\star {\cal E}^{(3)}$ in (\ref{deltaZ}) can now be reinterpreted as the BI of $F_7$, ${\cal Z}^{(8)}_{A_1 \dots A_8}=0$,  so that $\delta_\gamma{\cal Z}^{(5)}$ vanishes off-shell.   

Then, we proceed as follows: we begin by considering the most general transformation for $A_3$,  in terms of $E_{M}{}^{A}, A_{M_1 M_2 M_3}$ and $A_{M_1 \dots M_{6}}$, with arbitrary coefficients. From this, we first compute the transformation of the curvature $F_4$, then we impose the duality relation $F_7\to \star F_4$ and finally, we require agreement with the transformation obtained in the standard formulation (\ref{deltaF11}), which hopefully fixes all the coefficients in the ansatz for $\delta_\gamma A_3$.  Interestingly, this procedure leads to the unique solution
\begin{eqnarray}
\delta_\gamma A_{M_1 M_2 M_3} &=&
- \gamma_{M_1 M_2 M_3} 
-\frac{1}{12} (A\cdot\gamma) A_{M_1 M_2 M_3}       
+ \frac16 \gamma^{N_1 N_2 N_3} A_{M_1 M_2 M_3 N_1 N_2 N_3} \cr
&&+ \,  \frac34 \gamma^{N_1 N_2 N_3} A_{N_1 N_2 [M_1} A_{M_2 M_3] N_3}\;, \label{gammaA3}
\end{eqnarray}
such that the curvature $F_4$ transforms as
\begin{eqnarray}
\delta_\gamma  F_{MNPQ} &=&  - \frac{1}{3!}  A_{RST} \gamma^{RST} {F}_{MNPQ} 
- 2 \gamma^{RST}  A_{RS[M} {F}_{NPQ]T} 
\cr
&& -  4\; \partial_{[M}\gamma_{NPQ]} 
+ \frac{1}{3!} \gamma^{RST} F_{RSTMNPQ} \;,\label{gammaF4}
\end{eqnarray}
which reduces to (\ref{deltaF11}) after  implementing the duality relation $F_7\to \star F_4$, as expected.
Of course, $\delta_\gamma{\cal Z}^{(5)}=0$ is now trivially satisfied, as the variation of the curvature is defined through $\delta_\gamma A_3$.  

The question now is whether we can assign a $\gamma$ transformation to $A_6$. To answer that, we express  the transformation of $\star F_4$ as:
\begin{eqnarray}
\delta_\gamma(\star F_4)_{M_1 \dots M_7} &=& -(\star d\gamma)_{M_1 \dots M_7} - (\gamma\wedge F_{4})_{M_1 \dots M_7} -\frac{1}{3} (A\cdot \gamma) (\star F_{4})_{M_1 \dots M_7} \cr
&+& \frac72 \gamma^{N_1 N_2 N_3} (\star F_4)_{N_1[M_1 \dots M_6} A_{M_7] N_2 N_3}  \, .\label{deltaStarF4}
\end{eqnarray}
Then, guided by the experience in Type II supergravity, we propose that the transformation rule for the dual $F_7$ field is\footnote{We emphasize that this is a minimal choice, involving only the replacement $\star F_4\rightarrow F_7$. Replacing $F_4$ by any combination $aF_4 - b\star F_7$, with $a+b=1$, would be a non-minimal option, consistent with the duality relation.
}  
\begin{eqnarray}
\delta_\gamma F_{M_1 \dots M_7} &=& -(\star d\gamma)_{M_1 \dots M_7} - (\gamma\wedge F_{4})_{M_1 \dots M_7} -\frac{1}{3} (A\cdot \gamma) F_{M_1 \dots M_7} \cr
&+& \frac72 \gamma^{N_1 N_2 N_3} F_{N_1[M_1 \dots M_6} A_{M_7] N_2 N_3} \, . \label{gammaF7}
\end{eqnarray}

It turns out that $\delta_\gamma(dA_6)=\delta_\gamma (F_7 + \frac 1 2 A_3 \wedge F_4)$ is not exact, and then there is no chance to define the variation of  $A_6$. Indeed,   $\delta_\gamma(dA_6)$ is not even closed, as the transformation rules (\ref{gammaA3}), (\ref{gammaF4}) and (\ref{gammaF7}) yield:
\begin{eqnarray}
d\left[\delta_\gamma \left(F_7+ \frac12 A_3\wedge F_4\right)\right]= - d[\star (d \gamma)]+ A_3\  {\rm dependent\ terms}\neq 0\;. \label{deltad2A6}    
\end{eqnarray}
This conclusion is robust and independent of our (minimalist) definition of $\delta_\gamma F_7$, since the term $d[\star (d \gamma)]$ will survive with any choice.

 The problem we encountered in the previous section reapears, albeit at a higher level. While before we dealt with the set of fields $(E_M{}^A \, , \, F_{M N P Q})$  in the standard formulation, now we managed to undress the variation of the three-form potential at the expense of including its dual six-form. The set of fields has now raised to $(E_M{}^A \, , \, A_{M N P} \, , \, F_{M_1 \dots M_7})$, and their $\gamma$ variations are given by (\ref{gammaE}), (\ref{gammaA3}) and (\ref{gammaF7}), respectively.  Although the $\gamma$ transformation of $A_6$ cannot be defined at this level, we managed to move one step further.

An interesting opportunity arises from this extension. Notice that the variation of $G_{MN}$ in (\ref{gammag}) depends on $A_3$. If we wanted to explore the closure conditions involving $\gamma$ variations, we would have to face the problem that the transformation of $A_3$ is not accessible in the standard formulation. However, in this democratic improvement we can  evaluate the impact of $\gamma$ transformations  on the  closure of the symmetry algebra on the metric $[\delta_{1} \, , \, \delta_{2}] G_{M N}$, since we have the transformation (\ref{gammaA3})  at hand.  We find that closure is achieved as\begin{equation}
[\delta_{1} \, , \, \delta_{2}] G_{M N} = L_{\xi_{12}} G_{M N} + \delta_{\Gamma_{12}} G_{MN}\ ,
\end{equation}
 up to diffeomorphisms $\xi$ and global transformations with respect to a six-vector $\Gamma$. The brackets are given by  $\xi_{12}^M = [\xi_1 \, , \, \xi_2]+\frac 1 2 \gamma_{[1}^{MNP} \lambda_{2]NP}$ (where $\lambda_2$ is the gauge parameter for $A_3$), and 
\be
\Gamma_{12}^{M_1 \dots M_6} = \gamma_1^{[M_1 M_2 M_3} \gamma_2^{M_4 M_5 M_6]}
\ee
with 
\begin{eqnarray}
\delta_\Gamma G_{M N} &=& - \frac 1 3 \Gamma_{(M}{}^{M_1 \dots M_5} A_{N) M_1 \dots M_5} + \frac 1 {3^3} G_{M N} \Gamma^{M_1 \dots M_6} A_{M_1 \dots M_6} \nonumber\\ 
&&  + \frac 5 3 \Gamma_{(M}{}^{M_1 \dots M_5} A_{N) M_1 M_2} A_{M_3 M_4 M_5}  \ .
\end{eqnarray}
We should next evaluate the closure of the six-vector transformations on the metric. But  this is unattainable at this stage, because we lack the $\Gamma$ transformations of $A_3$ and $A_6$.

Unfortunately, we cannot  explore the closure conditions on $A_3$, since its $\gamma$ variation \eqref{gammaA3} depends on $A_6$. It seems that yet another field would be necessary. The situation is reminiscent of the conjecture that
 the infinite-dimensional Kac-Moody algebra $E_{11}$ is a symmetry of 11-dimensional supergravity  \cite{West,west2}. The main evidence for this conjecture is given
 by the observation that the level decompositions of $E_{11}$ with respect to its finite-dimensional
 subalgebras reproduce the same $p$-form representations as expected for maximal supergravity
 when formulated in a democratic way,  introducing for each $p$-form also its dual. In the 11-dimensional
 decomposition, the lowest levels of the theory include, beyond the metric, the three-form and the six-form at levels 0, 1 and 2, respectively, a field $h_{M_1\cdots M_{8},N}$ in a mixed-Young tableaux representation, which
 is interpreted as the dual of the graviton. In the next section we will see that the addition of a dual graviton allows to assess the $\gamma$ transformation of $A_6$.

Before exploring the effect of introducing a new field, we note that 
 one can construct a pseudo-action for the set of fields ($E_M{}^A, A_{MNP}, A_{M_1\cdots M_6}$) as
 \be
 S = \int d^{11}x \sqrt {-G} \left(R - \frac 1 {3\cdot 4!} F_{M_1\dots M_4} F^{M_1\dots M_4} - \frac 1 {6 \cdot 7!}  F_{M_1 \dots M_7} F^{M_1 \dots M_7}\right)\ ,
\ee
with $F_7$ defined  in (\ref{F7}), in the spirit of the democratic action of Type II, in the sense that  $A_6$ must be treated as an independent field,  there are no Chern-Simons terms  and  the duality relation $F_7 = \star F_4$  must be imposed as a constraint on the equations of motion. The field equation  for $A_6$ then becomes the BI $dF_4 = 0$.

There are other proposals for a reformulation of the standard action of 11-dimensional supergravity  containing both  $A_3$ and $A_6$ \cite{deal}-\cite{bdrh}. While all of them contain a topological term,  the duality relation is enforced on the field equations in \cite{deal} but it emerges  as the equation of motion of either a Lagrange multiplier in \cite{bbs,nishino} or an auxiliary gauge field in \cite{bdrh}.

\section{Towards the dual graviton }
 Motivated by the   level
 decomposition of $E_{11}$ \cite{West}, which includes  a mixed-Young tableaux representation interpreted as the dual of the graviton, 
 in this section we explore the possibility of defining the $\gamma$ transformation of the six-form potential by introducing new degrees of freedom related to the gravitational sector.  We discuss various challenges that arise in determining the BI of the dual graviton, and demonstrate how consistency with $\gamma$ symmetry, combined with certain expected symmetries of the dual graviton, ultimately determine its form up to a unique parameter, and provides a $\gamma$ transformation for $A_6$.

 The construction of a dual theory of gravity has been an elusive task for years. Even though linearized Einstein gravity in $D$-dimensions can be equivalently formulated
 in terms of a dual mixed Young tableaux field $h_{\mu_1\cdots\mu_{D-3},\nu}$ \cite{curt}-\cite{hull}, the no-go theorems of \cite{Bek,Bekaert}
 prove that there is no  manifestly Poincar\'e invariant, non-abelian, local deformation of the
  theory, and so there is no consistent non-abelian self-interaction of the dual graviton. Some proposals have been made  for a reformulation of (super-)gravity that contains the dual graviton
 and is valid at the non-linear level,  based on the introduction of extra gauge degrees of freedom and duality relations \cite{{bdrh},Boulanger:2008nd,Hohm}. However, in these extended theories, the gravitational sector does not mix with the $p$-form sector. Instead, in the $E_{11}$ conjecture, the symmetry
  transforms the fields of different spin into each other \cite{West}, allowing to evade   the no-go theorems  with the assumption that the field $h_{D-3,1}$ on its own does not correctly describe gravity at the non-linear level   \cite{Tumanov,west5}.

 The scope of this section is more modest. We just observe that pursuing the procedure implemented in the previous section, leads to the introduction of a new field related to gravity, which allows to define the $\gamma$ transformation of $A_6$. To see this, it is instructive to recap the systematic route that we tracked before, in order to find the $\gamma$ transformation of $A_3$. Starting with the set of fields ($E_M{}^A, F_{MNPQ}$), we applied the following steps:
\begin{enumerate}
\item We noticed that the $\gamma$ transformation of the BI for $F_4$ contained the equation of motion of $A_3$:
\be\delta_\gamma {\cal Z}^{(5)} \sim \star {\cal E}^{(3)} \ .\ee

\item We then imposed a duality relation $\star F_4 \to F_7$, which allowed to interpret $\star {\cal E}^{(3)}$ as the BI for $F_7$
\be
\star {\cal E}^{(3)} \to {\cal Z}^{(8)} \ ,
\ee
requiring the introduction of the potential $A_6$.

\item Finally, the transformation $\delta_\gamma A_3$ could be found in terms of $A_6$, by demanding that it reproduced the transformation of $F_4$, upon  imposition of the duality relation.
\end{enumerate}

 In the last step, we also obtained the transformation $\delta_\gamma F_7$, but this did not allow us to derive $\delta_\gamma A_6$. 
 To make progress, we now try to repeat the procedure described above, starting with the enlarged set of fields $(E_M{}^A , \, A_{MNP} \, , \, F_{M_1\cdots M_7})$. We will see that this requires the inclusion of the next level of the $E_{11}$ decomposition: the dual graviton.

Implementing step i. with this extended set of fields requires the $\gamma$ transformation of the BI for $F_7$. Hence, defining 
\begin{eqnarray}
{\cal Z}^{(8)}=\sqrt{-G} \; \left(dF_7+ \frac12 F_4\wedge F_4\right)\ ,    
\end{eqnarray}
we find
\begin{eqnarray}
\delta_\gamma {\cal Z}^{(8)}_{M_1 \dots M_8} &=&
-\frac12 (\star {\cal E}^{(2)})_{M_1 \dots M_8 PQ, N} \ \gamma^{NPQ}
+ \frac{1}{3!} \left({\cal Z}^{(5)}\wedge \gamma\right)_{M_1 \dots M_8}\cr 
&& - \frac{4}{9} (A\cdot \gamma) \ {\cal Z}^{(8)}_{M_1 \dots M_8} 
- {\cal Z}^{(8)}_{N[M_1 \dots M_7} A_{M_8]PQ} \gamma^{NPQ}\, ,\label{deltaZ8}
\end{eqnarray}
where $(\star {\cal E}^{(2)})_{M_1 \dots M_{10}, N} = 
\varepsilon_{P M_1 \dots M_{10}} {\cal E}^{(2)}_{N}{}^{P}$. Then, 
the upgrade of item i. can be stated as follows:
\begin{itemize}
\item[{\it i})]  We observe that the $\gamma$ transformation of the BI for $F_7$ contains
the dual of the gravitational equation of motion:
\be\delta_\gamma {\cal Z}^{(8)} \sim \star {\cal E}^{(2)} \ . \label{starE2}\ee

\end{itemize}

Now,  various questions emerge when trying to upgrade item ii. First, it requires a duality relation for the curvature of the gravitational field $E_M{}^A$. Different proposals for this duality can be found in the literature. The Hodge dual of the Riemann tensor, of the spin connection or of a combination of coefficients of anholonomy have been considered in  \cite{Tumanov},\cite{bdrh}-\cite{Hohm}.  Here, we note that a natural definition follows from the problematic term 
$(\star d\gamma)$ appearing in (\ref{gammaF7}). Actually, the equality 
\begin{eqnarray}
-(\star d\gamma)_{M_1 \dots M_7 } &=& -\frac{1}{4!}\varepsilon_{N_1 \dots N_4 M_1 \dots M_7} 4\nabla^{N_1}\gamma^{N_2 N_3 N_4}\cr     &=& \frac12 \varepsilon_{M_1 \dots M_7 N_1 \dots N_4} \omega_{P}{}^{AB} E^{N_1}{}_{A} E^{N_2}{}_{B} \gamma^{P N_3 N_4}\;,\label{stardgamma}
\end{eqnarray}
suggests the identification $\star\omega\rightarrow H$, where $\omega$ is the spin connection (the curvature of $E_M{}^A$), and $H$ is some curvature related to dual gravity as
\begin{eqnarray}
H_{A_1 \dots A_9,B}=-\frac{2}{9!}\varepsilon_{C D A_1 \dots A_9}\;  \omega_{B}{}^{CD}   \ , \ \ \  \ \omega_{A, B_1 B_2}=\frac{1}{4}\varepsilon^{C_1 \dots C_9}{}_{B_1 B_2} H_{C_1 \dots C_9, A}  \, ,  
\end{eqnarray}
in agreement with \cite{Tumanov}.

To proceed forward, we should  interpret $\star{\cal E}^{(2)}$ as the BI for $H_{9,1}$:
\be
\star{\cal E}^{(2)}\rightarrow {\cal Z}^{(10,1)}\, ,
\ee
and solve the equation ${\cal Z}^{(10,1)}=0$ in terms of a potential $h_{8 ,1}$: the  dual graviton.
But these steps face a number of ambiguities. On the one hand, 
one can dualize different expressions for the Ricci tensor  contained in ${\cal E}^{(2)}$  \eqref{eomeab}\footnote{$R_{AB}=2\partial_{[B}\omega_{C]A}{}^C-\omega_{CB}{}^D\omega_{DA}{}^C-\omega_C{}^{CD}\omega_{BAD}$ is not manifestly symmetric. Hence, one can dualize this expression, the equivalent one with $A\leftrightarrow B$ or a combination of both.}, and moreover, one can take diverse dualizations of the terms involving $\omega^2$ in $R_{AB}$\footnote{One can dualize $\omega^2$ as $\omega^2\to (\star H)^2$ or as $\omega^2\to\star H \omega$. The Levi-Civita tensor that is contracted with the spin connection in  $\star R\subset \star {\cal E}^{(2)}$ does not  yield $H_{9,1}$ directly. Instead, one must substitute $\omega$ in terms of $\star H$ and evaluate contractions between two $\varepsilon$-tensors.}.    These options lead to different gravitational BI, that are expected to   coincide only on-shell. 
On the other hand, the term $\epsilon_{BA_1 \ldots A_{10}} |F_4|^2$ contained in $\star {\cal E}^{(2)}_{A_1 \ldots A_{10} B}$ can be rewriten as
\be\epsilon_{BA_1 \ldots A_{10}} |F_4|^2=-\epsilon_{BA_1 \ldots A_{10}} |\star F_4|^2 = (\star F_4 \wedge F_4)_{BA_1 \ldots A_{10}}\, .\ee  In which  expression should one  replace $\star F_4\rightarrow F_7$? Furthermore, since $|F_4|^2$ is related to the scalar curvature $R$ on-shell, the relative coefficient between both of them is not completely fixed in ${\cal E}^{(2)}$.

Most probably, these ambiguities can be  resolved by including further global symmetries into the game (such as the six-vector symmetries discussed in the previous section). However, this is far beyond the scope of this work. Nevertheless, although we are unable at this stage to systematically reach the BI ${\cal Z}^{(10,1)}$, we can still try to find  the explicit dependence of $H_{9,1}$ on its potential $h_{8,1}$, and derive the $\gamma$ transformation of $A_6$.

Schematically, we  proceed as follows. We propose generic expressions for $H_{9,1}$ and $\delta_\gamma A_6$ in terms of $A_3, A_6$ and $h_{8,1}$, and then we demand that $\delta_\gamma F_7$ calculated from $\delta_\gamma A_6$ reproduces (\ref{gammaF7}) after imposing the duality relations $H \rightarrow \star \omega$ and $F_7 \rightarrow \star F_4$.

We consider the following ansatze for $H_{9,1}$ 
\begin{eqnarray}
H_{M_1 \dots M_{9},N} &=&  \alpha_1 \ \partial_{[M_1}h_{M_2 \dots M_9], N} 
+   \alpha_2 \ \partial_{N} h_{[M_1 \dots M_{8},M_9]} 
+   \alpha_3 \ \partial_{[M_1} h_{|N|M_2,\dots M_{8},M_9]} \cr 
&+& \alpha_4 \ A_{[M_1 M_2 M_3} \partial_{M_4}A_{M_5 M_6 M_7} A_{M_8 M_9]N} 
 +   \alpha_5 \ \partial_{[M_1} A_{M_2 \dots M_7} A_{M_8 M_9] N} \cr
&+&    \alpha_6 \ \partial_{M_1}A_{M_2 \dots M_6 |N|} A_{M_7 M_8 M_9} 
+   \alpha_7 \ \partial_{N}A_{[M_1 \dots M_6} A_{M_7 M_8 M_9]} \cr
&+&   \alpha_8 \ A_{N[M_1 \dots M_5} \partial_{M_6}A_{M_7 M_8 M_9]}
+   \alpha_9 \ A_{[M_1 \dots M_6} \partial_{|N|}A_{M_7 M_8 M_9]} \cr
&+&   \alpha_{10} A_{[M_1 \dots M_6} \partial_{M_7} A_{M_8 M_9]N} \ ,\label{H}
\end{eqnarray}
and for $\delta_\gamma A_6$ 
\begin{eqnarray}
\delta_\gamma A_{M_1 \dots M_6} &=&
a_1 \ h_{M_1 \dots M_6 N P, Q}\ \gamma^{NPQ} 
+ \ a_2 \ h_{NPQ [M_1 \dots M_5,M_6]} \ \gamma^{NPQ} 
+ \ a_3 \ (A_3\wedge \gamma)_{M_1 \dots M_6} \cr
&+& \left( \vphantom{\frac12}\right.  a_4\, A_{M_1 \dots M_6} A_{NPQ} 
+ a_5 \, A_{N[M_1 \dots M_5} A_{M_6]PQ} 
+ a_6 \, A_{NP[M_1 \dots M_4} A_{M_5 M_6]Q} \cr 
&+& a_7 \, A_{NPQ [ M_1 M_2 M_3}A_{M_4 M_5 M_6]}
+ a_8 \ A_{N[M_1 M_2} A_{M_3 M_4 M_5} A_{M_6]PQ} 
\left. \vphantom{\frac12}\right) \gamma^{NPQ}\; .\label{deltaA6Aux}
\end{eqnarray}
  Explicit dependence on the Levi-Civita tensor or terms involving contractions of $A_3$  and $A_6$  with themselves or with each other through a metric are not included, because they are absent in the ansatz (\ref{gammaF7}) and hence they are not expected to appear here. Including such terms in $H_{9,1}$ would also require them in (\ref{deltaA6Aux}), leading to a conflict with (\ref{gammaF7}).

Following \cite{west5}, we supplement these expressions with the conditions 
\begin{eqnarray}
h_{[M_1 \dots M_8,N]}=0\;,\qquad\qquad
H_{[M_1 \dots M_9,N]}=0  \;\; ,\label{Hconstraint}
\end{eqnarray}
which allows to eliminate the term proportional to $  \alpha_2$ in (\ref{H}) and to combine the terms with coefficients $\alpha_1$ and $  \alpha_3$ as well as the first two terms in equation (\ref{deltaA6Aux}). Therefore, we can set $  \alpha_2=  \alpha_3=a_2=0$, without any loss of generality. Furthermore, we can also set $a_1=1$, which is always possible through a renormalization of the field $h_{8,1}$. These restrictions, together with the imposition  (\ref{gammaF7}), lead to the refined ansatze
\begin{eqnarray}
H_{M_1 \dots M_{9},N} &=& 
-\frac{4}{8!} \ \partial_{[M_1}h_{M_2 \dots M_9], N} 
-\frac{1}{(3!)^3}  \ A_{[M_1 M_2 M_3} \partial_{M_4} A_{M_5 M_6 M_7} A_{M_8 M_9]N} \cr 
&-& \frac{1}{6!} \frac23\left(\vphantom{\frac12} \ \partial_{[M_1} A_{M_2 \dots M_7} A_{M_8 M_9] N} 
+  \partial_{M_1}A_{M_2 \dots M_6 |N|} A_{M_7 M_8 M_9} \right) \cr
&-&  \frac{2}{6!}\frac13\ \partial_{[M_1} \left( 3 \ e\ 
A_{M_2 \dots M_7} A_{M_8 M_9] N} - a_9 \ A_{M_2 \dots M_6 |N|} A_{M_7 M_8 M_9]}\right)\cr
&+&   \alpha_7 \ \partial_{N}A_{[M_1 \dots M_6} A_{M_7 M_8 M_9]} +    \alpha_9 \ A_{[M_1 \dots M_6} \partial_{|N|}A_{M_7 M_8 M_9]}\ ,\label{HFinal}
\end{eqnarray}
and
\begin{eqnarray}
\delta_\gamma A_{M_1 \dots M_6} &=&
 h_{M_1 \dots M_6 N P, Q}\ \gamma^{NPQ} 
- \ \frac12 \ (A_3\wedge \gamma)_{M_1 \dots M_6} \cr
&+& \frac56 \left(\vphantom{\frac12} 3\ A_{N[M_1 M_2} A_{M_3 M_4 M_5} A_{M_6]PQ} -\ 2 \ A_{NPQ [ M_1 M_2 M_3}A_{M_4 M_5 M_6]}
\right)\gamma^{NPQ}\cr
&+& a_4 \left( \vphantom{\frac12}\right.   A_{M_1 \dots M_6} A_{NPQ} 
+ 12  \ A_{N[M_1 \dots M_5} A_{M_6]PQ} 
+ 15  \ A_{NP[M_1 \dots M_4} A_{M_5 M_6]Q}
\left. \vphantom{\frac12}\right) \gamma^{NPQ}\cr
&+& a_9  \left( \vphantom{\frac12}\right. 
A_{N[M_1 \dots M_5} A_{M_6]PQ} 
+ 5 \ A_{NP[M_1 \dots M_4} A_{M_5 M_6]Q} \cr
&& \;\;\;\;\;\;\;\;\;\;\;\;\;\;\;\;\;\;\;\;\;\;\;\;
\;\;\;\;\;\;\;\;\;\;\;\;\;\;\;\;\;\;\;\;\;\;\;\;\;
+ \frac{10}{3} \ A_{NPQ [ M_1 M_2 M_3}A_{M_4 M_5 M_6]}
\left. \vphantom{\frac12}\right) \gamma^{NPQ}\ ,\label{deltaA6}
\end{eqnarray}
where we have introduced the parameter $a_9=\frac{3a_7+5}{10}$. 

There are a few additional simplifications that can be  considered. First, we can redefine the field variable $h_{8,1}$ to reduce the two parameters $a_4$ and $a_9$ into a single one $\alpha = 3a_4 + a_9 $. Specifically, we propose
\begin{eqnarray}
h_{M_1 \dots M_8,N} \to 
h_{M_1 \dots M_8,N} -\frac{a_9 \cdot 7 \cdot 4}{3}  \left(
A_{[M_1 \dots M_6} A_{M_7 M_8] N} + \ A_{[M_1 \dots M_5 |N|} A_{M_6 M_7 M_8]}\right)\, ,
\end{eqnarray}
which ensures that $h_{[M_1...M_8,N]} = 0$ remains valid if it was true initially. Finally, we can use (\ref{Hconstraint}), which further implies $  \alpha_7=  \alpha_9=-\frac{2 \cdot \alpha}{3\cdot 6!}$. The final expressions are then fixed, up to a unique parameter $\alpha$, as
\begin{eqnarray}
\delta_\gamma A_{M_1 \dots M_6} &=&
 h_{M_1 \dots M_6 N P, Q}\ \gamma^{NPQ} 
- \ \frac12 \ (A_3\wedge \gamma)_{M_1 \dots M_6} \label{deltaA6Final}\\
&+& \frac56 \left(\vphantom{\frac12} 3\ A_{N[M_1 M_2} A_{M_3 M_4 M_5} A_{M_6]PQ} -\ 2 \ A_{NPQ [ M_1 M_2 M_3}A_{M_4 M_5 M_6]}
\right)\gamma^{NPQ}\cr
&+& \frac{\alpha}{3}  \left( \vphantom{\frac12}\right.   A_{M_1 \dots M_6} A_{NPQ} 
+ 12  \ A_{N[M_1 \dots M_5} A_{M_6]PQ} 
+ 15  \ A_{NP[M_1 \dots M_4} A_{M_5 M_6]Q}
\left. \vphantom{\frac12}\right) \gamma^{NPQ} \ ,\nonumber
\end{eqnarray}
and
\begin{eqnarray}
H_{M_1 \dots M_{9},N} &=& 
-\frac{4}{8!} \ \partial_{[M_1} h_{M_2 \dots M_9], N} 
-\frac{1}{(3!)^3}  \ A_{[M_1 M_2 M_3} \partial_{M_4} A_{M_5 M_6 M_7} A_{M_8 M_9]N} \cr 
&-& \frac{1}{6!} \frac23\left(\vphantom{\frac12} \ \partial_{[M_1} A_{M_2 \dots M_7} A_{M_8 M_9] N} 
+  \partial_{[M_1}A_{M_2 \dots M_6 |N|} A_{M_7 M_8 M_9]} \right) \cr
&-& \alpha\ \frac{1}{6!}\frac23 \left[\vphantom{\frac12} \partial_{[M_1}\left(
A_{M_2 \dots M_7} A_{M_8 M_9] N}\right) + \partial_{N}\left( A_{[M_1 \dots M_6} A_{M_7 M_8 M_9]}\right)\right]\label{HFinal} \ .
\end{eqnarray}

In summary, we have found that the introduction of a new field $h_{8,1}$, dual to the gravitational degrees of freedom, allows to define $\delta_{\gamma} A_6$ in a way that remains compatible with $\delta_{\gamma} F_7$, ensuring that $\delta_{\gamma} {\cal Z}^{(8)}=0$ off-shell.  Even though it seems to be difficult to   derive of the Bianchi identity for $H_{9,1}$ from our bottom-up perspective, we expect that once it is understood, the $\alpha$-parameter can be ultimately fixed and the symmetries of the dual graviton (\ref{Hconstraint}) can be determined.

\section{Conclusions}

In this work, we have presented an 11-dimensional uplift of the tri-vector symmetry inherent to the exceptional groups arising from the compactification of maximal  supergravity on tori. This  was implemented through the application of an 11-dimensional constraint that assumes the existence of isometries. By uplifting the $\beta$ symmetry of Type II supergravity \cite{Beta}$-$\cite{TypeII}, we elaborated on the $\gamma$ transformations relevant to the metric and the curvature of the three-form. Our findings elucidate how the  equations of motion and Bianchi identities of the standard formulation of 11-dimensional supergravity can be structured into tri-vector multiplets. This implies that $\gamma$ transformations should be understood as an on-shell symmetry of  the theory, that can be employed as a solution generating technique.

A significant aspect of our analysis involved the derivation of the $\gamma$ transformation of the three-form potential, which required the introduction of a dual magnetic six-form. This addition, accompanied by a corresponding duality relation, is reminiscent of structures found in democratic formulations of supergravity. Additionally, we noticed that closure of the symmetry algebra indicates the presence of an 11-dimensional six-vector global symmetry. The six-form potential and its associated non-geometric six-vector symmetry represent the first meaningful extension of the standard 11-dimensional supergravity framework, in alignment with the $E_{11}$ construction \cite{West}.

Our results establish a bottom-up perspective on the $E_{11}$ formulation, offering direct insight into the conventional supergravity framework. We have further investigated the subsequent level of the $E_{11}$ decomposition and demonstrated that the $\gamma$ transformation of the curvature of the six-form yields a duality with the gravitational equations of motion. Thus, we have seen that advancing within this hierarchy  leads to the emergence of a dual graviton. We initiated the first steps in this intriguing direction, but a number of ambiguities turned up, which we expect to eventually clear up in future research.

The tri-vector $\gamma$ symmetry examined here promotes  the  symmetry 
 principles known as $\alpha$ \cite{Alpha} and  $\beta$ \cite{Beta} to a trilogy, based on the philosophy of extrapolating well defined symmetries of a given theory, to another theory connected to the original one through  dimensional reduction. This list is far from exhaustive, and it would be interesting to find new examples of this phenomenon, which has provided powerful ways of constraining interactions in supergravity and its extensions. 

Looking ahead, it might be worth to focus future research on exploring the relationships between the $\gamma$ symmetry examined here and the tri-vector deformations discussed in related literature \cite{Kulyabin:2022yls}-\cite{Barakin:2024rnz}. Moreover, a detailed examination of tri-vector symmetries in the context of higher-derivative corrections in 11-dimensional supergravity remains a promising avenue for exploration. Ultimately,  refining our formalism further, we hope to gain new insights into the dual graviton, which may enhance our understanding of the underlying symmetries in gravitational theories, and uncover connections with previous work \cite{west5},\cite{bdrh}-\cite{Hohm}.

\bigskip

{\bf Acknowledgements:} DM warmly thanks Chris Blair, Emanuel Malek and Edvard Musaev,  for an enlightening brainstroming during the initial stages of this project and to Edvard Musaev for comments on the manuscript. WB thanks Peter West for clarifications on his work. 
Support by Consejo Nacional de Investigaciones Cient\'ificas y T\'ecnicas (CONICET), Universidad de La Plata (UNLP) and Universidad de Buenos Aires (UBA) is gratefully acknowledged.

\begin{appendix}
\section{Notation and Definitions}\label{AppA}

We use  $\mu,\nu,\rho,\dots$ and  $a,b,c,\dots$ indices for space-time and tangent space coordinates in 10 dimensions, respectively. The infinitesimal Lorentz transformation of the vielbein is
\be
\delta_\Lambda e_\mu{}^a = e_\mu{}^b \Lambda_b{}^a \ \, .
\ee
The  spin connection 
\be
\omega_{c a b} = -e^\mu{}_c\left(\partial_\mu e_{\nu a}e^\nu{}_b -\Gamma^\rho_{\mu\nu} e_{\rho a} e^\nu{}_b\right)\, ,\qquad {\rm with }\ \ \ \ \  \Gamma^\rho_{\mu\nu}=\frac12 g^{\rho\sigma}\left(\partial_\mu g_{\sigma\nu}+\partial_\nu g_{\mu\sigma}-\partial_\sigma g_{\mu\nu}\right)\, ,
\ee
transforms as
\be
\delta_\Lambda \omega_{c a b} =  D_c \Lambda_{a b} + \omega_{d a b}  \Lambda^d{}_c  + 2 \omega_{c d [b} \Lambda^d{}_{a]}\, ,
\ee
and hence, it turns flat derivatives $D_a$ into covariant flat derivatives $\nabla_a$ as
\be
\nabla_a T_b{}^c = D_a T_b{}^c+ \omega_{a b}{}^d T_d{}^c- \omega_{ad}{}^cT_b{}^d \ , \ \ \ D_a = e^\mu{}_a \partial_\mu\, .
\ee 
The Christoffel connection $\Gamma^\rho_{\mu\nu}$ turns spacetime  partial into covariant derivatives as
\be
\nabla_\mu T_{\rho}{}^\sigma
=\partial_\mu T_{\rho}{}^\sigma-\Gamma_{\mu\rho}^\lambda T_\lambda{}^\sigma +\Gamma_{\mu\lambda}^\sigma T_\rho{}^\lambda\, .
\ee

The Riemann tensor 
\be
R^\mu{}_{\nu\rho\sigma}=\partial_\rho\Gamma^\mu_{\nu\sigma}-\partial_\sigma\Gamma^\mu_{\nu\rho}+\Gamma^\mu_{\rho\lambda}\Gamma^\lambda_{\nu\sigma}-\Gamma^\mu_{\sigma\lambda}\Gamma^\lambda_{\nu\rho}\, ,
\ee
with flat spacetime indices is defined as
\be
R_{a b c d} = 2 D_{[a}\omega_{ b] c d} + 2 \omega_{[a b]}{}^e \omega_{e c d} + 2 \omega_{[\underline{a} c}{}^e \omega_{\underline{b}] e d}\, .
\ee
While the symmetry $R_{a b c d} = R_{[ab] [cd]}$ is manifest, other symmetries of the Riemann tensor are hidden and  determine the Bianchi identities
\be
R_{a b c d} = R_{c d a b} \ , \ \ \  \ R_{[a b c] d} = 0 \\ , \ \ \  \ \nabla_{[a}R_{b c] d e} = 0  . \label{BI2}
\ee
The Ricci tensor and scalar curvature are given by the traces
\be
R_{a b} = R^c{}_{a c b} \ , \ \ \  R = R_a{}^a \ .  \ \ \ 
R_{[a b]} = 0\ . \label{BI3}
\ee

The same definitions apply in 11 dimensions, with the only difference being that the indices 
$M,N,P,\dots$ and $A,B,C,\dots$ now represent curved and tangent space indices, respectively.

\section{No-Go for off-shell $\gamma$-invariance in standard  11-dimensional supergravity}\label{AppNoGo}
We have seen that the standard formulation of 10-dimensional Type IIA supergravity is not off-shell $\beta$ invariant, which implies that the  standard formulation of 11-dimensional supergravity cannot be $\gamma$ invariant. 
In this Appendix we present a more fundamental proof of this statement, without evoking the RR sector of Type IIA. We  demonstrate that the circle reduction of the most general proposal for tri-vector transformations of the metric and the three-form potential is not compatible with the $\beta$ transformations  rules of the NSNS sector of Type IIA supergravity. 

We begin by considering the most general $\gamma$ transformations of the metric and the 3-form potential, which can be expressed as follows:
\begin{eqnarray}
\delta_\gamma G_{MN} = \delta_{\gamma^{(G)}} G_{MN} + \delta_{\gamma^{(\epsilon)}} G_{MN}\;,\;\;\;\;\;\;\;\;\;\;
\delta_\gamma A_{MNP} = \delta_{\gamma^{(G)}} A_{MNP} + \delta_{\gamma^{(\epsilon)}} A_{MNP}\;,
\end{eqnarray}
where we distinguish terms that involve the Levi-Civita tensor  and those that do not. Furthermore, we decompose the variation $\delta_\gamma$ into a sum, according to the number $(k)$ of 3-form potentials, i.e., $\delta_\gamma \to \sum_{k} \delta_\gamma^{(k)}$.
\begin{eqnarray}
\delta_{\gamma^{(G)}}G_{MN} &=& \sum_{k\geq0} \delta^{(2k+1)}_{\gamma^{(G)}} G_{MN}\;, \;\;\;\;\;\;\;
\delta_{\gamma^{(\epsilon)}}G_{MN} = \sum_{k\geq0} \delta^{(2k)}_{\gamma^{(\epsilon)}} G_{MN}\;, \cr
\delta_{\gamma^{(G)}}A_{MNP} &=& \sum_{k\geq0} \delta^{(2k)}_{\gamma^{(G)}} A_{MNP}\;, \;\;\;\;\;\;\;\;
\delta_{\gamma^{(\epsilon)}}A_{MNP} = \sum_{k\geq0} \delta^{(2k+1)}_{\gamma^{(\epsilon)}} A_{MNP}\;. 
\end{eqnarray}
The most general candidates for the lowest-order contributions  can be expressed as:
\begin{eqnarray}
\delta^{(1)}_{\gamma^{(G)}}G_{MN} &=& a_1 (\gamma\cdot A) G_{MN} + b_1 \gamma^{PQ}{}_{(M} A_{N)PQ} \;,\\  
\delta_{\gamma^{(G)}}^{(3)}G_{MN} &=& a_3 (\gamma\cdot A) A^2 G_{MN} + b_3 \gamma^{PQ(M} A_{N)PQ} A^2 
+ c_3 
(\gamma\cdot A) A^{PQ}{}_{M} A_{NPQ}\;, \\
\delta_{\gamma^{(G)}}^{(0)}A_{MNP} &=& a_0 \gamma_{MNP}\;,\label{deltaA}\\  
\delta_{\gamma^{(G)}}^{(2)}A_{MNP} &=& a_2 (\gamma\cdot A) A_{MNP} + b_2 \gamma^{RS}{}_{[M} A_{NP]}{}^{T} A_{RST}
+ c_2 \gamma^{R}{}_{[MN} A_{P]}{}^{ST} A_{RST} + d_2 \gamma_{MNP} A^2\;, \cr
&\dots& 
\end{eqnarray}
where $(\gamma\cdot A)=\gamma^{PQR} A_{PQR}$ and  $A^2=A^{PQR} A_{PQR}$.
The variations $\delta_{\gamma^{(\epsilon)}}$ necessarily involve higher-order contributions. In particular,  $\delta_{\gamma^{(\epsilon)}}^{(0)} G_{MN} = \delta_{\gamma^{(\epsilon)}}^{(2)} G_{MN} = 0$ and $\delta_{\gamma^{(\epsilon)}}^{(1)} A_{MNP} = \delta_{\epsilon}^{(3)} A_{MNP} = 0$, and the first nontrivial combinations (consistent with $\beta$ symmetry when truncating RR fields and $\gamma^{\mu\nu\rho}$) appear in $\delta_{\gamma^{(\epsilon)}}^{(4)} G_{MN}$ and $\delta_{\gamma^{(\epsilon)}}^{(5)} A_{MNP}$. 

The strategy to determine the transformation rules is to vary the action and require the cancellation of terms order by order in powers of the 3-form potential. Finally, we  impose that the circle reduction of the transformations reproduces the known $\beta$ transformations of the NSNS fields.

The 11-dimensional Lagrangian takes the form:
\begin{eqnarray}
{\cal L}_{11}={\cal L}^{(0)}+{\cal L}^{(2)}+{\cal L}^{(3)}\;,    
\end{eqnarray}
where ${\cal L}^{(0)}$ and ${\cal L}^{(2)}$ are respectively the Ricci scalar and the kinetic term of the 3-form potential up to the volume measure $\sqrt{|G|}$, while ${\cal L}^{(3)}$ is the Chern-Simon term $A_3\wedge F_4\wedge F_4$.

The general variation can thus be organized as:
\begin{eqnarray}
\delta_\gamma {\cal L}&=& \frac{\delta \left({\cal L}^{(0)}+ {\cal L}^{(2)}\right)}{\delta G_{MN}} \delta_{\gamma^{(G)}}G_{MN} + \frac{\delta {\cal L}^{(2)}}{\delta A_{MNP}} \delta_{\gamma^{(G)}}A_{MNP} + \frac{\delta {\cal L}^{(3)} }{\delta A_{MNP}} \delta_{\gamma^{(\epsilon)}}A_{MNP} \cr
&+&  \frac{\delta \left({\cal L}^{(0)}+ {\cal L}^{(2)}\right)}{\delta G_{MN}} \delta_{\gamma^{(\epsilon)}}G_{MN} + \frac{\delta  {\cal L}^{(2)}}{\delta A_{MNP}} \delta_{\gamma^{(\epsilon)}}A_{MNP}  + \frac{\delta {\cal L}^{(3)} }{\delta A_{MNP}} \delta_{\gamma^{(G)}}A_{MNP}\; ,
\label{deltaL}
\end{eqnarray}
where we have used $\frac{\delta  {\cal L}^{(0)}}{\delta A_{MNP}}= \frac{\delta {\cal L}^{(3)}}{\delta G_{MN}}=0$. Notice that the first line has an odd number of potentials $A$, while the second line has an even number of them, implying that each line must vanish independently.

Compatibility with $\delta_\beta b_{\mu\nu}= -\beta_{\mu\nu} + ...$ requires $a_0\neq 0$ in (\ref{deltaA}). This leads to a non-vanishing quadratic contribution in (\ref{deltaL})
\begin{eqnarray}
\frac{\delta {\cal L}^{(3)} }{\delta A_{MNP}} \delta_{\gamma^{(G)}}^{(0)} A_{MNP} \sim \gamma \wedge F \wedge F    \sim A^2 . \label{deltaL3}
\end{eqnarray}
This term cannot be cancelled either by the first two terms in the second line of  (\ref{deltaL}), since
\begin{eqnarray}
\frac{\delta \left({\cal L}^{(0)}+ {\cal L}^{(2)}\right)}{\delta G_{MN}} \delta_{\gamma^{(\epsilon)}}G_{MN} + \frac{\delta {\cal L}^{(2)}}{\delta A_{MNP}} \delta_{\gamma^{(\epsilon)}}A_{MNP}    \sim A^4\;,
\end{eqnarray}
(recall that $\delta_{\gamma^{(\epsilon)}}G_{MN}\sim A^4$ and $\delta_{\gamma^{(\epsilon)}}A_{MNP}\sim A^5$) or by
\begin{eqnarray}
\frac{\delta {\cal L}^{(3)} }{\delta A_{MNP}} \delta_{\gamma^{(G)}}^{(k)} A_{MNP}  \sim A^{k+2} \;,
\end{eqnarray}
with $k\geq2$.

In conclusion, we have found that there is no off-shell uplift of $\beta$ symmetry to the standard formulation of 11-dimensional supergravity. Note that $\gamma$ in (\ref{deltaL3}) is given by 
\begin{eqnarray}
\gamma_{MNP} = G_{MQ} G_{NR} G_{PS} \gamma^{QRS}\;,
\end{eqnarray}
which is not constant. Therefore, $\gamma \wedge F \wedge F$ is not a total derivative, indicating that neither the Lagrangian nor the action is $\gamma$ invariant.

\end{appendix}


\begin{thebibliography}{99} 

\bibitem{Beta} W.~H.~Baron, D.~Marques and C.~A.~Nunez,
``\ensuremath{\beta} Symmetry of Supergravity,''
Phys. Rev. Lett. \textbf{130} (2023) no.6, 061601
[arXiv:2209.02079 [hep-th]].

\bibitem{beta2} W.~H.~Baron, D.~Marques and C.~A.~Nunez,
``Exploring the $\beta$ symmetry of supergravity,''
JHEP \textbf{12} (2023), 006
[arXiv:2307.02537 [hep-th]].

\bibitem{TypeII}
W.~H.~Baron and N.~A.~Yazbek,
``\ensuremath{\beta} symmetry in type II supergravities,''
JHEP \textbf{03} (2024), 146
[arXiv:2312.15061 [hep-th]].

\bibitem{Heterotic}
W.~H.~Baron, C.~A.~Nunez and J.~A.~Rodriguez,
``$\beta$ symmetry of heterotic supergravity,''
[arXiv:2410.17067 [hep-th]].

\bibitem{West} P.~C.~West,
``E(11) and M theory,''
Class. Quant. Grav. \textbf{18} (2001), 4443-4460
[arXiv:hep-th/0104081 [hep-th]]. 

\bibitem{west2} F.~Riccioni and P.~C.~West,
``Dual fields
 and E(11),''
Phys. Lett. B \textbf{645} (2007), 286-292
[arXiv:hep-th/0612001 [hep-th]].

\bibitem{west3} P.~West,
``$E_{11}$, generalised space-time and IIA string theory,''
Phys. Lett. B \textbf{696} (2011), 403-409
[arXiv:1009.2624 [hep-th]].


\bibitem{Tumanov}
A.~G.~Tumanov and P.~West,
``E11 and the non-linear dual graviton,''
Phys. Lett. B \textbf{779} (2018), 479-484
[arXiv:1710.11031 [hep-th]].



\bibitem{west5}
K.~Glennon and P.~West,
``The non-linear dual gravity equation of motion in eleven dimensions,''
Phys. Lett. B \textbf{809} (2020), 135714
[arXiv:2006.02383 [hep-th]].


\bibitem{bko} E.~Bergshoeff, R.~Kallosh, T.~Ortin, D.~Roest and A.~Van Proeyen,
``New formulations of D = 10 supersymmetry and D8 - O8 domain walls,''
Class. Quant. Grav. \textbf{18} (2001), 3359-3382
[arXiv:hep-th/0103233 [hep-th]].

\bibitem{nicolai}
H.~Nicolai, P.~K.~Townsend and P.~van Nieuwenhuizen,
``COMMENTS ON ELEVEN-DIMENSIONAL SUPERGRAVITY,''
Lett. Nuovo Cim. \textbf{30} (1981), 315

\bibitem{DAuria}
R.~D'Auria and P.~Fre,
``Geometric Supergravity in d = 11 and Its Hidden Supergroup,''
Nucl. Phys. B \textbf{201} (1982), 101-140

\bibitem{townsend}
P.~K.~Townsend,
``P-Brane Democracy,''
[arXiv:hep-th/9507048 [hep-th]].

\bibitem{West:2011mm}
P.~West,
``Generalised geometry, eleven dimensions and E11,''
JHEP \textbf{02} (2012), 018
[arXiv:1111.1642 [hep-th]].

\bibitem{deal} S.~P.~de Alwis,
``Coupling of branes and normalization of effective actions in string / M theory,''
Phys. Rev. D \textbf{56} (1997), 7963-7977
[arXiv:hep-th/9705139 [hep-th]].



\bibitem{bbs} I.~A.~Bandos, N.~Berkovits and D.~P.~Sorokin,
``Duality symmetric eleven-dimensional supergravity and its coupling to M-branes,''
Nucl. Phys. B \textbf{522} (1998), 214-233
[arXiv:hep-th/9711055 [hep-th]].

\bibitem{nishino}
H.~Nishino,
``Alternative formulation for duality symmetric eleven-dimensional supergravity coupled to super M five-brane,''
Mod. Phys. Lett. A \textbf{14} (1999), 977-992
[arXiv:hep-th/9802009 [hep-th]].

\bibitem{bdrh} E.~A.~Bergshoeff, M.~de Roo and O.~Hohm,
``Can dual gravity be reconciled with E11?,''
Phys. Lett. B \textbf{675} (2009), 371-376
[arXiv:0903.4384 [hep-th]].

\bibitem{curt}
T.~Curtright,
``GENERALIZED GAUGE FIELDS,''
Phys. Lett. B \textbf{165} (1985), 304-308

\bibitem{Boulanger:2022arw}
N.~Boulanger, P.~P.~Cook, J.~A.~O'Connor and P.~West,
``Higher dualisations of linearised gravity and the $ {A}_1^{+++} $ algebra,''
JHEP \textbf{12} (2022), 152
[arXiv:2208.11501 [hep-th]].

\bibitem{Hull:2000zn}
C.~M.~Hull,
``Strongly coupled gravity and duality,''
Nucl. Phys. B \textbf{583} (2000), 237-259
[arXiv:hep-th/0004195 [hep-th]].

\bibitem{Hull:2001iu}
C.~M.~Hull,
``Duality in gravity and higher spin gauge fields,''
JHEP \textbf{09} (2001), 027
[arXiv:hep-th/0107149 [hep-th]].

\bibitem{hull}
C.~M.~Hull,
``Magnetic charges for the graviton,''
JHEP \textbf{05} (2024), 257
[arXiv:2310.18441 [hep-th]].

\bibitem{Bek}
X.~Bekaert, N.~Boulanger and M.~Henneaux,
``Consistent deformations of dual formulations of linearized gravity: A No go result,''
Phys. Rev. D \textbf{67} (2003), 044010
[arXiv:hep-th/0210278 [hep-th]].

\bibitem{Bekaert}
X.~Bekaert, N.~Boulanger and S.~Cnockaert,
``No self-interaction for two-column massless fields,''
J. Math. Phys. \textbf{46} (2005), 012303
[arXiv:hep-th/0407102 [hep-th]].


\bibitem{Boulanger:2008nd}
N.~Boulanger and O.~Hohm,
``Non-linear parent action and dual gravity,''
Phys. Rev. D \textbf{78} (2008), 064027
[arXiv:0806.2775 [hep-th]].

\bibitem{Hohm}
O.~Hohm and H.~Samtleben,
``The dual graviton in duality covariant theories,''
Fortsch. Phys. \textbf{67} (2019) no.5, 1900021
[arXiv:1807.07150 [hep-th]].

\bibitem{Alpha}
M.~Ciafardini, D.~Marques, C.~A.~N\'u\~nez and A.~P.~Grau,
``Hidden symmetries from extra dimensions,''
JHEP \textbf{02} (2025), 072
doi:10.1007/JHEP02(2025)072
[arXiv:2410.07325 [hep-th]].

\bibitem{Kulyabin:2022yls}
A.~Kulyabin and E.~T.~Musaev,
``SUSY and Tri-Vector Deformations,''
Symmetry \textbf{14} (2022) no.12, 2525
[arXiv:2210.14788 [hep-th]].

\bibitem{Bakhmatov:2022lin}
I.~Bakhmatov, A.~\c{C}atal-\"Ozer, N.~S.~Deger, K.~Gubarev and E.~T.~Musaev,
``Generalized 11D supergravity equations from tri-vector deformations,''
Eur. Phys. J. C \textbf{83} (2023) no.1, 37
[arXiv:2209.01423 [hep-th]].

\bibitem{Musaev:2023own}
E.~T.~Musaev and T.~Petrov,
``Tri-vector deformations on compact isometries,''
Eur. Phys. J. C \textbf{83} (2023) no.5, 399
[arXiv:2302.08749 [hep-th]].

\bibitem{Barakin:2024rnz}
S.~Barakin, K.~Gubarev and E.~T.~Musaev,
``Tri-vector deformations with external fluxes,''
Eur. Phys. J. C \textbf{84} (2024) no.12, 1312
[arXiv:2410.01629 [hep-th]].

\end{thebibliography}
\end{document}